# *ElecTra* Code: Full-Band Electronic Transport Properties of Materials

*Patrizio Graziosi*[1-2], *Zhen Li*[2], *and Neophytos Neophytou*[2]

[1] CNR – ISMN, Bologna, Italy

[2] School of Engineering, University of Warwick, UK

## Abstract

This paper introduces ElecTra, an open-source code which solves the linearized Boltzmann transport equation in the relaxation time approximation for charge carriers in a full-band electronic structure of arbitrary complexity, including their energy, momentum, and band-index dependence. ElecTra stands for 'ELECtronic TRAnsport' and computes the electronic and thermoelectric transport coefficients electrical conductivity, Seebeck coefficient, electronic thermal conductivity, and mobility, for semiconductor materials, for both unipolar and bipolar (small bandgap) materials. The code uses computed full-bands and relevant scattering parameters as inputs and considers single crystal materials in 3D and 2D. The present version of the code (v1) considers: i) elastic scattering with acoustic phonons and inelastic scattering with non-polar optical phonons in the deformation potential approximation, ii) inelastic scattering with polar phonons, iii) scattering with ionized dopants, and iv) alloy scattering. The user is given the option of intra- and inter-band scattering considerations. The simulation output also includes relevant relaxation times and mean-free-paths. The transport quantities are computed as a function of Fermi level position, doping density, and temperature. ElecTra can interface with any DFT code which saves the electronic structure in the '.bxsf' format. In this paper ElecTra is validated against ideal electronic transport situations of known analytical solutions, existing codes employing the constant relaxation time approximation, as well as experimentally well-assessed materials such as Si, Ge, SiGe, and GaAs.

# 1. Introduction

Electronic and thermoelectric transport simulations in novel complex bandstructure materials is an essential aspect of understanding material properties and optimizing them towards relevant applications **[1-4]**. A crucial element of this is the extraction of the scattering relaxation times that determine electronic transport, and their energy-, momentum- and band-dependence complexities **[5-11]**. In the absence of efficient and reliable tools to account for these, the majority of thermoelectric studies, for example, smear all these dependencies into a single number, typically $\tau = 10$ fs, referred to as the constant relaxation time approximation (CRTA) **[12, 13]**. State of the art first principles methods account for all scattering complexities by calculating billions of the matrix elements that contribute to the scattering rates **[14-20]**, but this makes them computationally extremely expensive, especially for 3D systems, **[14, 20]** inflexible in accounting for all major scattering mechanisms, and thus rarely used for such studies. Computationally inexpensive methods based on single parabolic bands **[21]** represent a first step to gain some accuracy beyond the CRTA, but appear limited to simulate complex bandstructures where full-band approaches seem more suitable.

The *ElecTra* simulator addresses the challenge of electronic and thermoelectric transport in complex bandstructure materials by solving the linearized Boltzmann transport equation (BTE) for charge transport in the relaxation time approximation. By using deformation potential theory and wave-vector and energy-dependent momentum relaxation times, *ElecTra* computes transport in a reliable and computationally attractive way, constituting the middle ground between the CRTA and fully *ab initio* scattering rate calculations. *ElecTra* considers the first-order solution of the BTE, **[22, 23]** and computes the charge transport coefficients by considering charge carrier scattering with phonons, ionized dopants, and alloy scattering, **[5, 9, 22, 23]** including bipolar effects.**[24, 25]** *ElecTra* takes as input the electronic bandstructure and scattering parameters, forms constant energy surfaces and scattering rate expressions, and returns the charge transport coefficients electrical conductivity ($\sigma$), Seebeck coefficient (*S*), thermoelectric power factor (PF = $\sigma S^2$), and electronic part of the thermal conductivity ($\kappa_e$) as function of Fermi level ($E_F$) position and temperature (*T*). Output results for the CRTA and constant mean-free-path considerations are also included. Finally, special attention is placed on user-friendliness, with well-defined input/output (I/O) files, GUI interfaces, and a detailed manual that accompanies the code for execution instructions.

We underline here that *ElecTra* offers the functionality of charge transport calculation beyond the constant relaxation time approximation by: i) considering the full energy/momentum/band dependence of the scattering rates, ii) distinction between intra- and inter-band transitions, including the ionized impurity scattering, and iii) bipolar transport by considering simultaneously the valence and conduction bands. Moreover, this is performed in a full-band approach and using anisotropic scattering rates, as detailed by Eqs. (7a), (7b), (7c), (7d), (7e), (7f) below, with the aim of an optimal trade-off between overall computational burden and accuracy. Indeed, a first step to relax the CRTA is to use energy dependent isotropic scattering rates in analytical equations and use the effective mass approximation. We have done this in the past to validate *ElecTra* and to gain conceptual insight in the physics of bipolar transport by using either parabolic and non-parabolic (Kane's model) bands. [**24**, **25**] This is very similar to the approach implemented in the TOSSPB code. [**21**] Further, another code, AMSET, employs full-band electronic structure coupled to momentum relaxation rates, in a similar manner to *ElecTra*. However, it merges all e-ph coupling in one parameter, the elastic acoustic phonon deformation potential, evaluated from unit cell expansion. [**18**] This is quite convenient to avoid ab initio matrix element calculation, with the drawback of not considering the fine details of the different e-ph scattering events separately. Other recent codes like EPA, [**15**] PERTURBO,[**19**] EPIC-STAR [**20**] compute matrix elements using ab-initio Density Functional Perturbation Theory (DFPT) and then perform different treatments to reduce the computational complexity, as for example averaging a few matrix elements over the whole BZ or evaluating the scattering rates along specific BZ paths. *ElecTra* uses deformation potentials to construct expressions for different phonon scattering processes for non-polar phonons, and in addition includes anisotropic scattering with polar phonons and ionized dopants. The deformation potentials can be computed ab initio using only a few matrix elements as we describe in Ref. [**26**]. At present, the deformation potentials must be entered as input information by the user. For the general material, a method to extract those from DFT is described in Ref. [**26**], and we plan to integrate such methods with the *ElecTra* code in the future.

The manuscript is organized as follows: In Section 2 we describe the theory behind the linearized BTE formalism used. In Section 3 we describe the way that we map the electronic structure onto constant-energy surfaces, as well as the method validation with analytically known solutions and other existing codes. In Section 4 we describe the calculation of the scattering rates. Finally, Section 5 presents illustrative example cases for comparison of *ElecTra* results and outputs to those of existing codes and known semiconductors. The paper is accompanied with three Appendices. Appendix A shows the process and results for 2D materials. Appendix B contains additional computational technicalities while Appendix C points to the supplementary material with examples of graphical user interfaces (GUI) for using the code as an app, as well as code examples to execute the code using text files and scripts instead of app GUIs.

## 2. Linearized Boltzmann Transport Equation (BTE) formalism

The transport (and TE) coefficients are computed using the transport distribution function (TDF) within the Linearized Boltzmann Transport equation as rank-2 tensors in the form [7, 9, 23]:

$$\sigma_{ij(E_F,T)} = q_0^2 \int_E \Xi_{ij}(E) \left(-\frac{\partial f_0}{\partial E}\right) dE, \tag{1a}$$

$$S_{ij(E_F,T)} = \frac{q_0 k_B}{\sigma_{ij}} \int_E \Xi_{ij}(E) \left(-\frac{\partial f_0}{\partial E}\right) \frac{E-E_F}{k_B T} dE, \tag{1b}$$

$$\kappa_{e_{ij}} = \frac{1}{T} \int_E \Xi_{ij}(E) \left(-\frac{\partial f_0}{\partial E}\right)(E-E_F)^2 dE - \sigma S^2 T \tag{1c}$$

where $\Xi_{ij}(E)$ is the Transport Distribution Function (TDF) defined below in Eq. (2), $E_F$, $T$, $q_0$, $k_B$, and $f_0$, are the Fermi level, absolute temperature, electronic charge, Boltzmann constant, and equilibrium Fermi distribution, respectively.

The TDF is expressed as a surface integral over the constant energy surfaces, $\mathfrak{L}_E^n$, for each band, and then summed over the bands, as [7, 22, 23]:

$$\Xi_{ij(E,E_F,T)} = \frac{s}{(2\pi)^3} \sum_{\mathbf{k},n}^{\mathfrak{L}_E^n} v_{i(\mathbf{k},n)} v_{j(\mathbf{k},n)} \tau_{i(\mathbf{k},n,E_F,T)} \frac{dA_{\mathbf{k}_{\mathfrak{L}_E^n}}}{|\vec{v}_{(\mathbf{k},n,E)}|} \tag{2}$$

where $\mathbf{k}_{\mathfrak{L}_E^n}$ represents a state on the surface $\mathfrak{L}_E^n$ and $dA_{\mathbf{k}_{\mathfrak{L}_E^n}}$ is its corresponding surface area element, computed as explained below in Section 3 [22]. $v_{i(\mathbf{k},n,E)}$ is the *i*-component of the band velocity of the transport state, $\tau_{i(\mathbf{k},n,E)}$ is its momentum relaxation time (combining the relaxation times of each scattering mechanism using Matthiessen's rule), $\frac{dA_{(\mathbf{k},n,E)}}{|\vec{v}_{(\mathbf{k},n,E)}|}$ is its density-of-states (DOS) [22], and *s* is the spin degeneracy. In the code *s* = 2 is used when the two spin sub-bands are degenerate, i.e. the material is non-ferromagnetic, but only one spin is resolved in the bandstrucutre, thus each band needs to account for both spins. In this case the user must enter that the bands are not spin-resolved, which is the case when spin-orbit coupling (SOC) is not considered or, for a non-ferromagnetic material, when SOC is considered in DFT, but the identical bands are removed from the full computed bandstructure. This step can be done when the bandstructure in .bxsf format is imported. Note that in Eq. (2) terms such as the velocity, can be uniquely defined by the *k*-point and the band index. We chose to let the energy in the triad (*k*,*n*,*E*) to specifically emphasize that all these quantities are also energy dependent.

he relaxation times for each individual scattering mechanism are combined following the Matthiessen's rule for each $\mathbf{k}_{\mathfrak{L}_E^n}$ state, to compute the comprehensive TE coefficients. In addition, TE coefficients are computed also for each considered scattering mechanism separately. Also, the overall energy-dependent relaxation time $\tau$ and mean-free-path (mfp) $\lambda$ are computed as well, both per-band, per scattering mechanism, and overall for all mechanisms. These are computed as:

$$\lambda_{i(E,E_\text{F},T)} = \frac{\sum_{k,n}^{\vartheta_E^n} |v_{i(k,n,E)} \tau_{i(k,n,E,E_\text{F},T)}| DOS_{(k,n,E)}}{\sum_{k,n}^{\vartheta_E^n} DOS_{(k,n,E)}} \qquad (3a)$$

$$\tau_{i(E,E_\text{F},T)} = \frac{\sum_{k,n}^{\vartheta_E^n} \tau_{i(k,n,E,E_\text{F},T)} DOS_{(k,n,E)}}{\sum_{k,n}^{\vartheta_E^n} DOS_{(k,n,E)}} \qquad (3b).$$

### 3. Electronic structure quantities

The workflow of *ElecTra* is shown in **Figure 1**. The electronic structure is entered as an input and it consists of a four-dimensional matrix (three-dimensional matrix for a 2D system) where the first three indexes are the space coordinates and the fourth is the band index. Because DFT codes usually use different formats, *ElecTra's* interface can take as input a .bxsf format file [**27**, **28**] enabling it to interface with any DFT code that provides the electronic structure data in this format. The code also checks the 'completeness' of the entered input instructions related for example to scattering mechanisms, Fermi levels, temperature ranges, and others as detailed in Appendix C (e.g. whether for each phonon process, its deformation potentials and required parameters are supplied). The code then starts the construction of the constant energy surfaces that will later form the density of states (DOS) in energy by mapping the $E(\boldsymbol{k})$ bandstructure into a $\boldsymbol{k}(E)$ one. At this level, the code computes also band-related properties such as the DOS and the band velocities, separately for each band, as well as comprehensive. Then, the code checks for the consistency between the scattering parameters and the input instructions and computes the scattering related quantities and the transport coefficients.

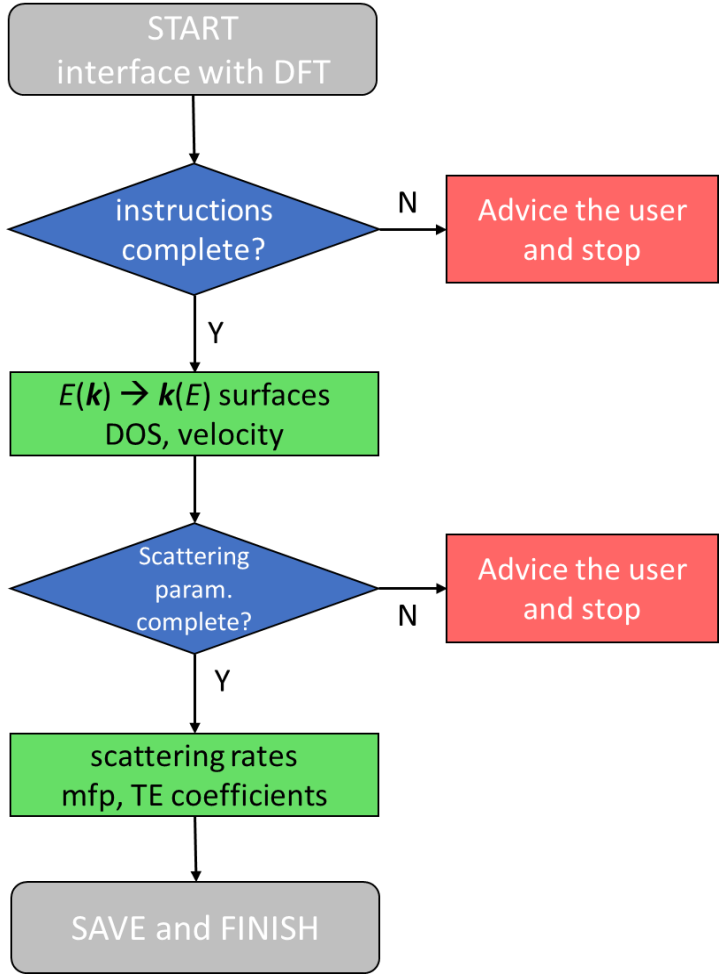

**Figure 1**: *ElecTra* workflow. At the beginning ElecTra interfaces with the output of the DFT codes to obtain the bandstructure; then, after a check of the entered calculation instructions, the constant energy surfaces are built. After this, the entered scattering parameters are checked, the scattering rates are computed, the energy dependent quantities are composed, and the TDF is integrated to obtain the TE coefficients. A file containing all the data is finally saved.

To map the $E(\boldsymbol{k})$ into a $\boldsymbol{k}(E)$, the 3D mesh in the $k$-space is scanned, the mesh elements crossed by the constant energy surfaces are identified and the coordinates of the points on these surfaces are computed. *ElecTra* offers two possibilities: i) A triangulation performed locally on the $k$-space mesh elements which are crossed by the surface of the constant energy of interest, by dividing it into six tetrahedra as shown in Figure 2a. Each of them can be crossed by the constant energy surface under consideration in two ways as shown in Figure 2b-c. ii) An easier approximate method which uses sampling of the nearest neighbour points on the $k$-mesh. Although the latter is an approximation because it detects only the points along the edges of the $k$-mesh elements, it is around 15 to 30 times faster, but without noticeable penalty in the results compared to the triangulation. The discrepancies in terms of density of states (DOS) are for isotropic bandstructures at the level of the numerical noise, while for anisotropic bands the differences increase from negligible at the band edge to ~ 1% around 0.2 eV from the band edge and to ~ 4% around 0.5 eV from the band edge. Details for the two methods are provided below.

Once a mesh element is identified, the three components of the band velocity $v_{i(\mathbf{k},n,E)}$ are computed with the contragradient method, suitable for the arbitrary symmetry of the BZ and its 3D $k$-space mesh. [29] For the coordinates of the points, the $E(\mathbf{k})$ is approximated to be linear between two points "v1" and "v2" of the $k$-mesh, which are taken as the vertexes of the element in **Figure 2b**, or the dots on the corners in **Figure 2d**. Thus, a $\mathbf{k}_i$-point at energy $E_i$ between the vertexes v1 and v2 is selected as [30]:

$$\mathbf{k}_i = \mathbf{k}_{v1} + (\mathbf{k}_{v2} - \mathbf{k}_{v1})\frac{E_i - E_{v1}}{E_{v2} - E_{v1}} \tag{4}.$$

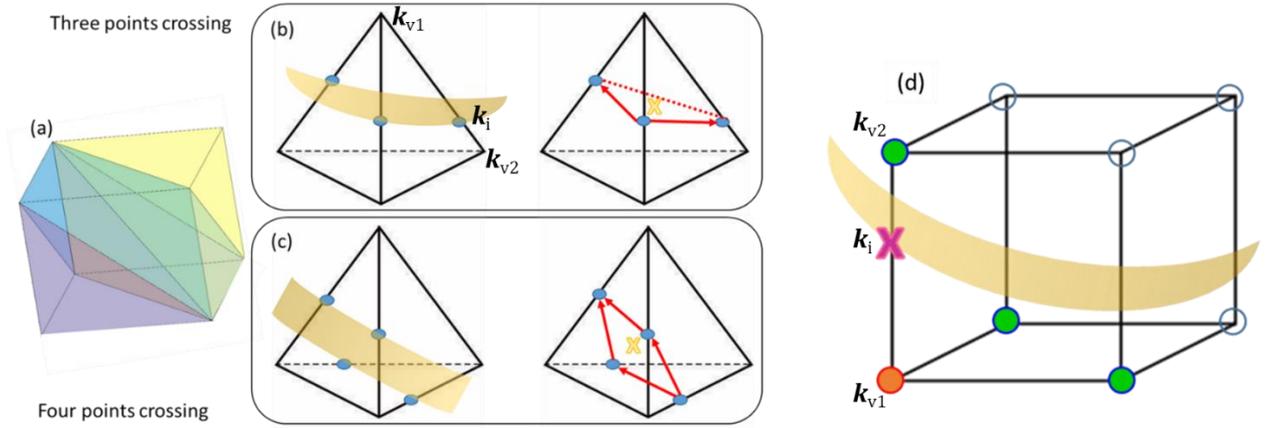

**Figure 2:** (a) Triangulation of a 3D mesh element. (b) and (c) show the cases where the constant energy surface of interest crosses a tetrahedron in three points (b) or four points (c). A part of the energy surface is depicted in yellow and the crossing points in blue. The surface element will be the area of the triangle in (b) or the parallelogram in (c) defined by the red arrows. Its associated element coordinates are represented by the coordinates of the barycentre shown by the yellow cross. (d) Conceptual scheme of the nearest-neighbours sampling. A $\mathbf{k}$-point in a 3D $k$-mesh element (orange) with some of its nearest neighbours (in green). When a constant energy surface (yellow) crosses the edge between the point under consideration and its neighbour, the crossing point (purple) is selected.

*Triangulation:* When the triangulation method is adopted, each tetrahedron will have three (**Figure 2b**), or four (**Figure 2c**), crossing points depending on how the energy surface crosses it. These points then define a surface whose area is computed with Heron's formula as $A = \sqrt{s_p(s_p - l_1)(s_p - l_2)(s_p - l_3)}$, where $A$ is the triangle area, $s_p$ is the semi-perimeter and $l_i$ are the triangle sides. When the tetrahedron has four points, the surface element is twice the average of the areas of the four defined triangles. Such surfaces are the surface area elements $dA_{(\mathbf{k},n,E)}$ defining the $\mathbf{k}$-state-dependent DOS, $\frac{dA_{(\mathbf{k},n,E)}}{|\vec{v}_{(\mathbf{k},n,E)}|}$, and used in the scattering rate calculations as well as the energy integrations. To obtain the $\mathbf{k}$-state coordinates, which is important for anisotropic scattering rates, for which the exchange vector is needed, the code finds the barycentre of the surface element, and labels this as the **k**-point associated to the given surface element.

*<u>Nearest-neighbour sampling method:</u>* When the nearest-neighbour sampling method is chosen, the code scans all the **k**-points on the energy surface and checks for their nearest neighbours along the edges of the **k**-mesh. If the **k**-point of interest and a neighbour have energies one above and one below the energy value of interest, the code selects a new **k**-point along the edge of the *k*-mesh element (**Figure 2d**), as in Eq. (4). This new point becomes the **k**-point at the energy of interest for the scattering rate evaluation. In this way the code does not resolve a constant energy surface element but acquires a collection of points on the energy surface of interest. Then, the code assigns an effective d$A_k$ surface element area value to each point to allow the extraction of the density of states associated with that **k**-point. For this, all the points on the constant energy surfaces are grouped as a cloud of points. The space in the neighbourhood of each **k**-point is explored to detect its neighbours on the surface. This is done in a radius of $1.25\sqrt[3]{V^*}$. Here $V^*$ is the effective volume associated with the **k**-mesh element computed as the average of the absolute values of each volume $V_e$ used in the contragradient method, i.e. $V_e = \boldsymbol{k}_{lm} \cdot (\boldsymbol{k}_{ln} \times \boldsymbol{k}_{mn})$, $\boldsymbol{k}_{lm} = (\boldsymbol{k}_l - \boldsymbol{k}_m)$ and *l*, *m*, *n*, are the vertexes of the considered mesh element. $\sqrt[3]{V^*}$ is regarded as the effective distance between adjacent points in the *k*-mesh used in the bandstructure calculation if the mesh is cubic. Then, the code calculates the average distance between the given point and its detected neighbours, $<\Delta k>$. The surface element associated to the **k**-point is approximated by a circle of radius half the average distance to the neighbouring points, i.e., $dA_{\boldsymbol{k}} = \pi \left(\frac{<\Delta k>}{2}\right)^2$. Note that the assumption of a circular region and the value of 1.25 is determined empirically here to provide the best map between the DOS of this method and the DOS of the triangulation method, and works adequately for 3D and 2D, but only if the sampling distance between **k**-points is uniform in all three directions. When the mesh sampling is not uniform we recommend to use the triangulation instead. Essentially, it indicates that the effective radius of a point to its neighbours needs to be somewhat more than its half distance in order to include points that are placed in the diagonal direction of the grid in relation to the considered point.

For each band in the electronic structure, the energy-dependent DOS is then calculated as:

$$\text{DOS}(E,n) = \oiint_{\Omega_E^n} \frac{dA_{(k,n,E)}}{|\vec{v}_{(k,n,E)}|} = \frac{s}{(2\pi)^3} \sum_{k_{E,n}} \frac{dA_{(k,n,E)}}{|\vec{v}_{(k,n,E)}|} \tag{5}$$

where *s* is the spin degeneracy taken as 1 or 2, and $\vec{v}_{(\boldsymbol{k},n,E)}$ is the band velocity. The comprehensive DOS(*E*) is the sum of the DOS of all individual bands and the comprehensive velocity $v(E)$ is the average of the state velocity $v(E,n) = \langle|\vec{v}_{(\boldsymbol{k},n,E)}|\rangle_{\boldsymbol{k}}$.

The implementation of these concepts in *ElecTra* are shown in **Figure 3** for three example cases: (i) parabolic bands, where the analytical solution is known, (ii) the valence band of the half-Heusler

TiCoSb, and (iii) the conduction band of Mg$_3$Sb$_2$. **Figure 3a-c** shows an example of constant energy surfaces together with 1D projections of the bandstructures, with the magenta lines indicating the energy of the surfaces. In **Figure 3d** the DOS of an isotropic parabolic band with mass equal to the rest mass of the electron, $m_0$, is shown, depicting in blue the analytical solution $DOS(E) = \frac{\sqrt{2}}{\pi^2 \hbar^3} m^{3/2} \sqrt{E}$, in orange the values computed by *ElecTra* with the nearest-neighbour sampling method and in green with the triangulation method. In the same sub-figure we also show the cases for an anisotropic non-parabolic band, following the same colouring scheme (bottom lines). For this case the three masses along the three coordinate directions are 1, 0.5, 0.1, in units of $m_0$, and the non-parabolicity coefficient is $\alpha = 0.5/q_0$ eV$^{-1}$, with $q_0$ being the electron charge. The analytical DOS in this case is $DOS(E) = \frac{\sqrt{2}}{\pi^2 \hbar^3} m_D^{3/2}(1 + 2\alpha E)\sqrt{E(1 + \alpha E)}$ with $m_D = \sqrt[3]{m_x m_y m_z}$. **Figure 3e** reports the band velocity for these two cases, with the isotropic parabolic band corresponding to the bottom lines, following the analytical solution $v(E) = \sqrt{2E/m}$. The anisotropic non-parabolic band case corresponds to the top lines, with analytical solution $v(E) = \frac{\sqrt{2E(1+\alpha E)/m_c}}{\sqrt{1+4\alpha E(1+\alpha E)}}$ with $m_c = \frac{3}{\sum_i 1/m_i}$. Excellent agreement between analytical and numerical calculations is observed. **Figure 3f** shows the total DOS for the TiCoSb valence band computed with Quantum Espresso [**31**, **32**] in blue, and with *ElecTra* in orange and green using nearest-neighbour sampling and triangulation, respectively. An excellent agreement between *ElecTra* and Quantum Espresso is observed. The corresponding band velocity is shown in **Figure 3g**. The comparison between *ElecTra* and BoltzTraP [**12**] is shown in **Figure 3h** for the Mg$_3$Sb$_2$ conduction band, whereas the corresponding band velocity is reported in **Figure 3i**. Again, excellent agreement between *ElecTra* and BoltzTraP is observed. Some minor differences are likely because *ElecTra* considers constant energy surfaces and surface integrals instead of fixed **k**-points and volume integrals.

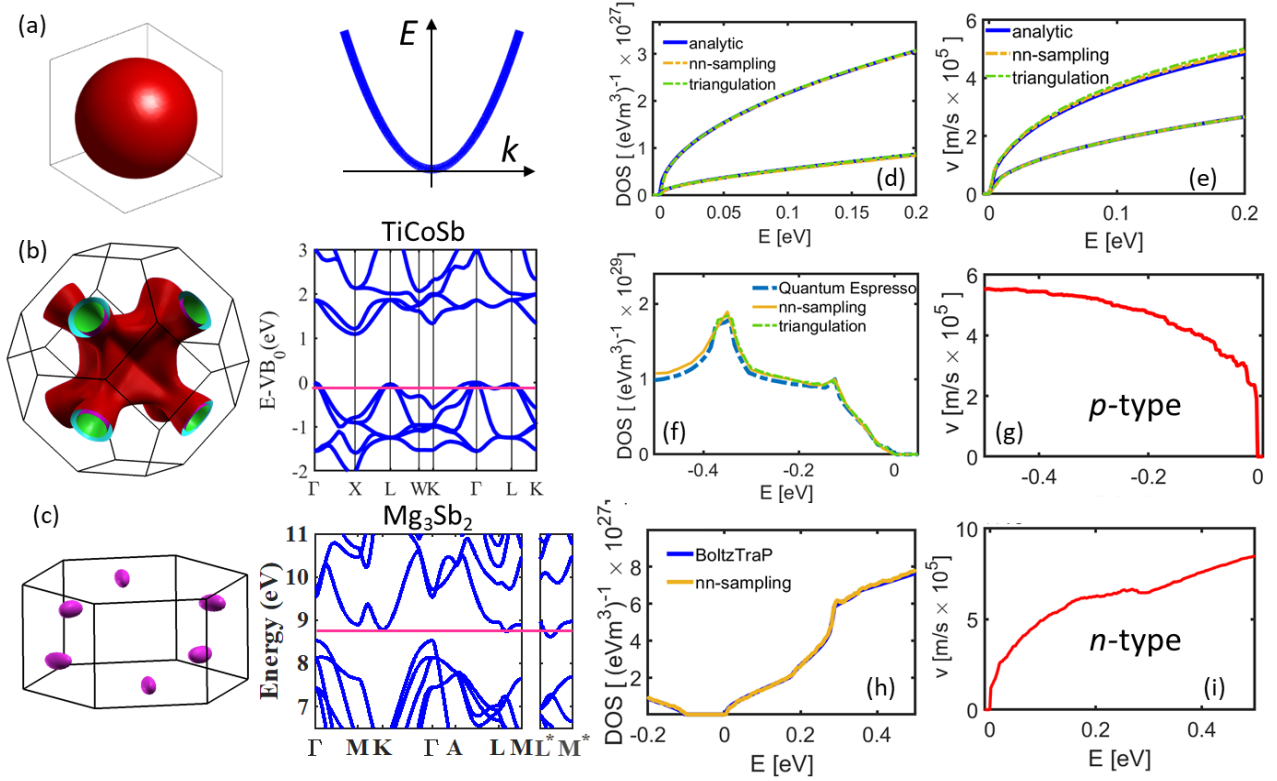

**Figure 3**: (a) Constant energy surface (sphere) for an isotropic parabolic band. (b) Constant energy surface for the valence band of TiCoSb, 0.12 eV below the valence band edge. (c) Constant energy surface for the Mg$_3$Sb$_2$ conduction band, 0.1 eV above the edge. (d) 3D DOS for isotropic parabolic band ($m^* = m_0$ – top lines), case (a), and anisotropic non-parabolic band ($m^*_{x,y,z} = 1, 0.5, 0.1\ m_0$ and $\alpha$ = 0.5/q$_0$ eV$^{-1}$ – bottom lines); comparison of the known analytical solution with the one numerically computed with *ElecTra*, by using the two sampling schemes for the BZ. (e) Band velocity for the case in (d), top lines for anisotropic non-parabolic band and bottom lines for isotropic parabolic. (f) DOS for the TiCoSb (Zincblende structure) valence band, comparison of the results from Quantum Espresso with those from ElecTra. (g) Band velocity for the TiCoSb valence band. (h) DOS for the Mg$_3$Sb$_2$ conduction band, comparison between the results from BoltzTraP and *ElecTra*. (i) Band velocity for the Mg$_3$Sb$_2$ conduction band.

## 4. Carrier scattering and transport quantities

For each transport state (***k***,*n*,*E*) and each scattering mechanism $m_s$, the corresponding momentum relaxation time $\tau^{(m_s)}_{i(k,n,E)}$ is defined as:

$$\frac{1}{\tau^{(m_s)}_{i(k,n,E)}} = \frac{1}{(2\pi)^3} \sum_{k'} |S^{(m_s)}_{k,k'}| \left(1 - \frac{v_{i(k')}}{v_{i(k,n,E)}}\right) \qquad (6)$$

where the sum runs over all the allowed final states ***k'*** of the same carrier spin, i.e. *ElecTra* will not allow scattering between them. [22, 23] $|S_{k,k'}|$ is the transition rate between the initial ***k*** and final ***k'*** states, computed as detailed by Eqs. (7a), (7b), (7c), (7d), (7e), (7f) below for the different mechanisms. The $\left(1 - \frac{v_{ij(k')}}{v_{ij(k,n,E)}}\right)$ term is an approximation for the momentum relaxation time, [33-36] which is used to solve the BTE in the closed form, as commonly done in the literature when computing the transport coefficients. [**14**, **22**, **35**, **36**] *ElecTra* computes the scattering rates using Fermi's Golden Rule for the different scattering mechanisms as (for 3D materials, see Appendix for the 2D versions) [**22**, **23**]:

$$\left|S_{k,k'}^{(ADP)}\right| = 2\frac{\pi}{\hbar}D_{ADP}^2\frac{k_BT}{\rho v_S^2}g_{k'} \tag{7a}$$

$$\left|S_{k,k'}^{(ODP)}\right| = \frac{\pi D_{ODP}^2}{\rho\omega}\left(N_\omega + \frac{1}{2}\mp\frac{1}{2}\right)g_{k'} \tag{7b}$$

$$\left|S_{k,k'}^{(IVS)}\right| = \frac{\pi D_{IVS}^2}{\rho\omega}\left(N_\omega + \frac{1}{2}\mp\frac{1}{2}\right)g_{k'} \tag{7c}$$

$$\left|S_{k,k'}^{(POP)}\right| = \frac{\pi q_0^2\omega}{|k-k'|^2\varepsilon_0}\left(\frac{1}{k_\infty}-\frac{1}{k_s}\right)\left(N_{\omega,BE}+\frac{1}{2}\mp\frac{1}{2}\right)g_{k'} \tag{7d}$$

$$\left|S_{k,k'}^{(IIS)}\right| = \frac{2\pi}{\hbar}\frac{Z^2q_0^4}{k_s^2\varepsilon_0^2}\frac{N_{imp}}{\left(|k-k'|^2+\frac{1}{L_D^2}\right)^2}g_{k'} \tag{7e}$$

$$\left|S_{k,k'}^{(Alloy)}\right| = \frac{2\pi}{\hbar}\Omega_c x(1-x)G_{k-k'}\Delta E_G^2 g_{k'} \tag{7f}$$

Above, ADP stands for 'Acoustic Deformation Potential' and represents the scattering between charge carriers and acoustic phonons. ODP stands for 'Optical Deformation Potential' and describes the charge carrier inelastic scattering with non-polar optical phonons. Both can can be chosen to be both intra- and/or inter-*band* (in subsequent versions of the code we will allow for the choice of intra- versus inter-*valley* scattering as well). IVS stands for 'Inter-Valley Scattering' and it is specific for the inelastic inter-valley scattering. POP stands for 'Polar Optical Phonon' and describes the inelastic/anisotropic scattering of charge carriers with polar phonons, which here is treated as both intra- and inter-band [**23**]. *ElecTra* allows different phonon frequencies for all these inelastic processes separately, for example, for each non-polar and polar phonon branches, different frequencies can be used . IIS stands for 'Ionized Impurity Scattering' and describes the elastic scattering rate due to ionized dopants, for which the user can choose both intra- and/or inter-band nature for transitions. "Alloy" represents the alloy scattering due to intrinsic disorder in alloys or solid solutions and is both intra- and inter-band. [**33**] *k* and *k'* are the wave vectors of the initial and final states. "-" and "+" in Eqs. (7b)–(7d) indicate the phonon absorption and emission processes, respectively. Examples for these types of transition processes are depicted in **Figure 4a**. We allow for the directional dependence of the momentum scattering rates because in an anisotropic band and /or under the influence of anisotropic scattering mechanisms, the rate at which the carrier's momentum relaxes, will depend on the carrier's initial momentum and the distribution of momenta of the final states.

The variables that appear in Eqs. (7a), (7b), (7c), (7d), (7e), (7f) are as follows: $D_{ADP}$, $D_{ODP}$, $D_{IVS}$ are the deformation potentials for the ADP, ODP, and IVS mechanisms. $\rho$ is the mass density, $v_s$ the sound velocity, $\omega$ the dominant frequency of optical phonons, considered as constant over the whole reciprocal unit cell, which has been validated to be a satisfactory approximation, [15] and $N_\omega$ is the phonon Bose-Einstein statistical distribution; $\varepsilon_0$ the vacuum dielectric constant, $k_s$ and $k_\infty$ the static and high frequency relative permittivity, $Z$ the electric charge of the ionized impurity considered, and $N_{imp}$ is the density of the ionized impurities. $g_{\boldsymbol{k}'} = \frac{dA_{(\boldsymbol{k}',n,E)}}{|\vec{v}_{(\boldsymbol{k}',n,E)}|}$ is the single-spin DOS of the final scattering state. $L_D = \sqrt{\frac{k_s \varepsilon_0}{e} \left(\frac{\partial n}{\partial E_F}\right)^{-1}}$ is the generalized screening lendegth with $E_F$ being the Fermi level and $n$ the carrier density. [23, 9] $\Omega_c$ is the volume of the primitive cell, $x$ the fraction of one of the alloy elements, and $\Delta E_G$ the difference between the energy gap of the two constituent materials that form the alloy. [23, 32] The $G$ function is the form factor of a hard sphere. [33] Rigorous studies show that the $\Delta E_G$ can be substituted by an effective scattering potential that can be calibrated to fit experimental values. [33] However, *ElecTra* uses the $\Delta E_G$ term, since in general such scattering potentials are unknown.

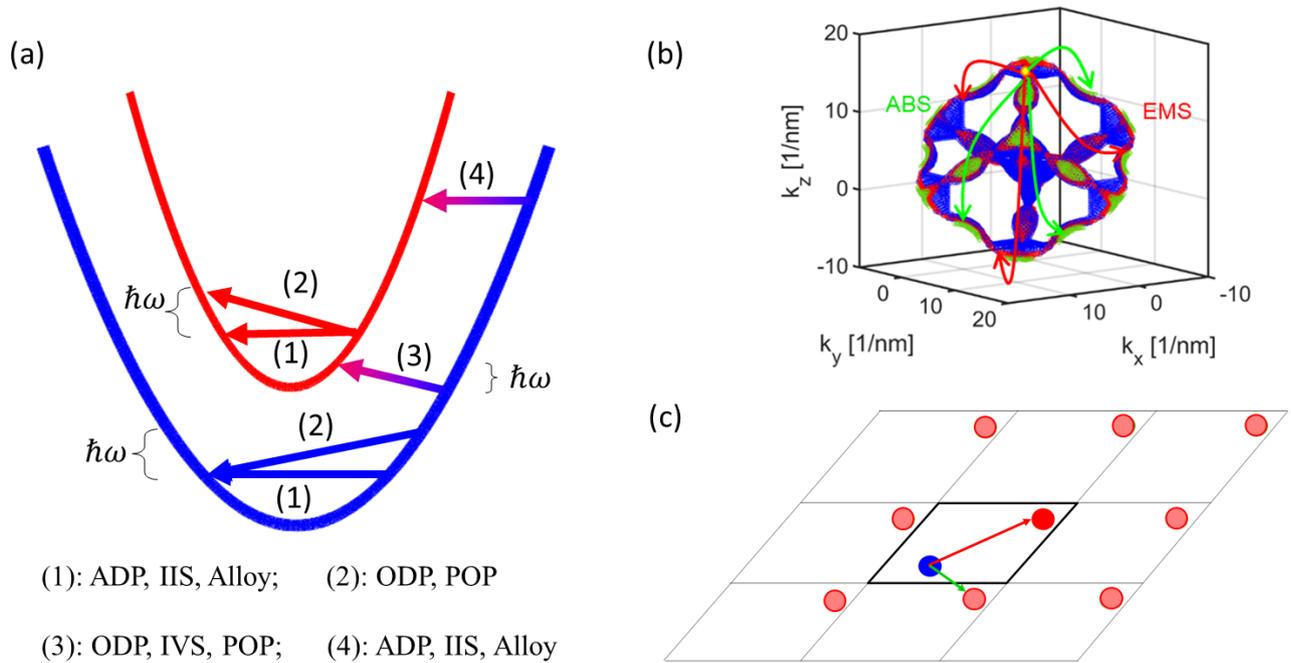

**Figure 4**: (a) Schematic of two bands with the four types of allowed transitions: (1) elastic intra-band; (2) inelastic intra-band, (3) inelastic inter-band, (4) elastic inter-band. $\hbar\omega$ is the energy of the absorbed or emitted phonon. (b) The same iso-energy surface as in **Figure 3b**, in blue, with transport state representation – each point is a transport state. Two other iso-energy surfaces (green/red) from another valence band and at different energies are depicted with relative arrows to signify absorption/emission transitions, indicated as ABS and EMS. To compute the tensor components of the TE coefficients, *ElecTra* expresses the surfaces, from DFT calculation on the BZ, in Cartesian axes. (c) 2D schematic of the reciprocal unit cell used in the calculation, with bold edges, and its equivalent cells around it. The initial $\boldsymbol{k}$-point is in blue and the final $\boldsymbol{k}$-point in red. The equivalent final $\boldsymbol{k}$-points are shown in fainted red, obtained by translating the red point by one reciprocal lattice vector in all possible directions. In the anisotropic POP and IIS scattering mechanisms, all the equivalent $\boldsymbol{k}$-points are explored, and the final $\boldsymbol{k}$-point considered is the one closest to the initial $\boldsymbol{k}$-point. This is used for the |$\boldsymbol{k}$-$\boldsymbol{k}$'| term and for charge screening (IIS and POP).

The current version of *ElecTra* (v1) is calibrated towards the predictive modelling of electronic and thermoelectric properties of materials and it is validated to room and higher temperatures. It considers that all dopants are ionized and does not capture the carrier freeze out region. Note that the equipartition theorem at the basis of Eq. (7a) does not hold at low temperatures either. Screening in the electron-phonon scattering processes is implemented only for the POP mechanism, where it is believed to play a major role, [**37-40**] for which the user has the choice to include the screening factor $\frac{1}{\left(1+\frac{1}{|k-k'|^2}\frac{1}{L_D^2}\right)^2}$ in the calculation. [**23, 41**] The screening in the carrier-phonon scattering event has different effect on transport depending on the bandstructure details. It has little effect for light bands with small ellipsoids and small DOS (larger $L_D$) such as in GaAs, the difference in mobility is below 5 % and concentrated in the low doping regime, and greater effect for bands with large constant energy surfaces and large DOS (smaller $L_D$) such as the valence bands of half-Heuslers, where the difference in the mobility can be ~ 50 % and extended in a wider doping interval. However, it substantially increases the computational complexity (as the scattering rates become Fermi level dependent), so its consideration is left to the user's choice.

The scattering rates and the transport coefficients are computed along orthogonal Cartesian space directions *x*, *y, z*. Consequently, the constant energy surfaces are expressed in Cartesian coordinates on orthogonal axes instead of unit cell axes, and the reciprocal unit cell is used instead of the Brillouin Zone. The surfaces in **Figure 3b**, represented as a collection of transport states in Cartesian coordinates, are depicted in **Figure 4b**. The process of carrier scattering is depicted by arrows between points (i.e. transport states).

A further important point is that the POP and IIS scattering strengths depend on the momentum exchange vector, i.e. the distance in the *k*-space between the initial and final states $|\boldsymbol{k} - \boldsymbol{k}'|^2$. To compute this momentum exchange vector, the simulator takes into consideration every final state's $\boldsymbol{k}$-point position in all neighbouring reciprocal unit cells, and then uses the minimum distance from the initial point in the exchange vector. For this, as shown in a 2D schematic in **Figure 4c**, the final point (red bullet), is shifted by the reciprocal lattice vectors in all possible directions, creating the fainted red bullet points. Here in **Figure 4c** the central cell with bold edges represents the cell used in the simulations and the other cells are the equivalent ones. Then the code considers the closest distant to the initial blue bullet point. This is necessary, because for example two $\boldsymbol{k}$-points that are located at opposite edges of the reciprocal unit cell are actually very near if the equivalent point in the nearest neighbour cell is considered. **Figure 4c** shows a 2D schematic of this, indicating that the physical scattering vector is the green one, and not the red one. In the case of an anisotropic dielectric constant or deformation potentials, at present we consider averaging of the quantities in the different directions, and are in the process of evaluating at what extent the anisotropy is important.

*Illustrative examples for transport quantities – case of parabolic bands:* We now show some illustrative examples of the energy dependent quantities computed by *ElecTra*. We consider the relaxation times $\tau$, mean-free-paths $\lambda$, and transport distribution functions $\Xi$, and first compare parabolic bands (with known analytical solutions). We then consider the valence band of TiCoSb as a case study. In **Figures 5a-d** we show the relaxation time and mean-free-path (mfp) computed for an isotropic parabolic band with $m = m_0$, under different scattering mechanisms as detailed in the figure. In addition to the scattering parameters detailed in the caption, we assumed a mass density of 9 g/cm$^3$, speed of sound 3 km/s, and static and high-frequency dielectric constants of 12 and 10. We consider room temperature $T = 300$ K. The behaviour is the typical one described in the literature [**22**]. In **Figure 5a** the relaxation time due to ADP decreases with energy due to the increasing DOS. The same generally holds for ODP with the additional consideration that at the band edge only phonon absorption is possible. When phonon emission becomes possible, the relaxation time drops down as expected. **Figure 5a** also shows the relaxation time for scattering with ionized impurity scattering at a dopant density of $2\times10^{19}$cm$^{-3}$, which corresponds to a Fermi level at the band edge. It also shows the total relaxation time by combining ADP, ODP and IIS according to Matthiessen's rule. In **Figure 5b** we show the mean-free-path, $\lambda$, for three of the situations of **Figure 5a**, for the parabolic bands. Namely, the mfps are: i) a roughly constant under ADP, ii) initially increasing under ODP when only absorption is possible, then decreasing and settling to a nearly constant value, and iii) following an increasing trend under IIS, which is the strongest for carriers at the band edge, such that $\lambda$ tends to gets closer to the phonon-limited value at higher energies. We show POP separately in **Figures 5c-d**, specifically differentiating the role of screening which can be included for this scattering mechanism. **Figures 5c** and **5d** show the relaxation time and mean-free-path, respectively, for the case where carrier screening is *not* included in the POP scattering calculation (blue lines), and when it *is* included (magenta lines). The three magenta lines show how these quantities vary for different carrier densities as indicated in **Figure 5d**. For non-degenerate conditions the effect is negligible, while it becomes sizeable at degenerate conditions by weakening the scattering strength by a factor of around 2 to 3.

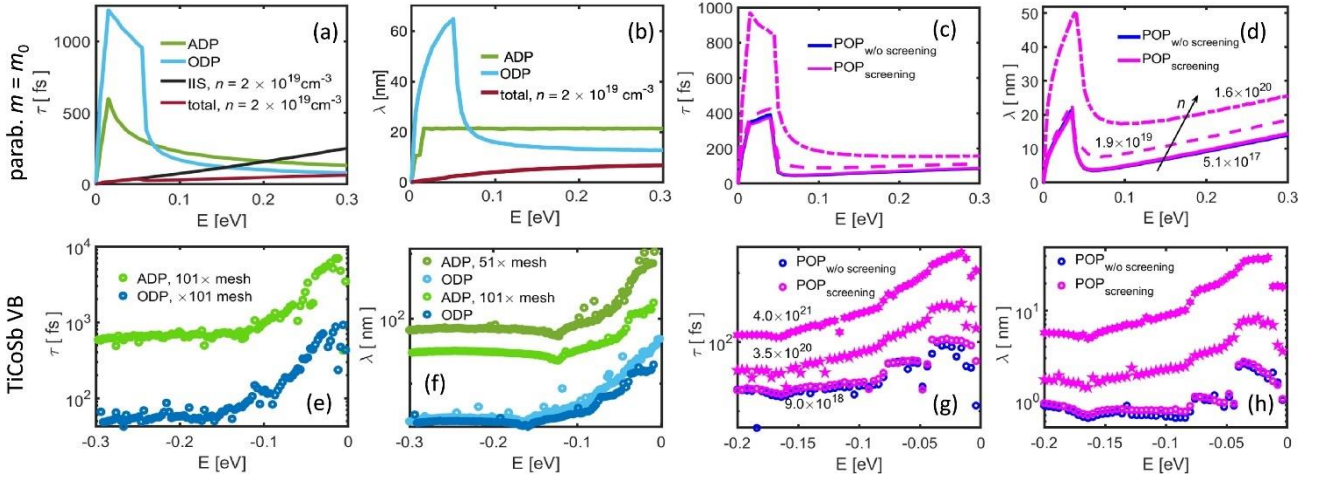

**Figure 5**: *ElecTra* simulation examples for: (a) Scattering times and (b) mean-free-paths for ADP (10 eV) and ODP (5×10$^{10}$ eV/m, phonon energy 50 meV) mechanisms in an isotropic parabolic band ($m = m_0$). (c) and (d) are the same as (a) and (b) for the POP mechanisms (phonon energy 40 meV), without (blue) and with (magenta) charge screening. In the latter case, the rates depend on the doping level. (e) Scattering times and (f) mean-free-paths for TiCoSb valence band for the ADP and ODP mechanisms. In (f) two *k*-mesh densities are compared. (g)-(h), same as (c)-(d) but for the TiCoSb valence band.

*Illustrative examples for realistic materials – the case of TiCoSb:* We now show the cases of relaxation times and mean-free-paths for the half-Heusler TiCoSb in **Figure 5e-h**, with the scattering parameters used taken from ref. [**9**]. Illustrative bandstructure figures are shown in **Figure 3b**. We start with ADP and ODP in **Figure 5e-f**. Note that the valence bands are in the negative energy direction and the band edge at 0eV, so the lines should be read from right to left. For both cases we have the rise of τ and λ at the band edge and reduction further into the bands. In **Figure 5f** we also compare the effect of the mesh size, by employing two regular meshes of 51 points and 101 points along each ***k***-space coordinate direction, respectively. A finer mesh, despite taking longer time, strongly reduces the numerical noise. The computation time is ~ 25 min. and increases to ~ 9 hrs when the mesh increases from 51 to 101 ***k***-points meshes, when *ElecTra* is parallelized on a 12 CPUs desktop PC.

In **Figures 5g-h**, we show the relaxation time and mean-free-path due to POP with and without screening. The blue lines correspond to the cases without screening, whereas the magenta lines for the cases of three carrier densities as above in **Figures 5c-d**. For TiCoSb, the dielectric constant values assumed in the POP calculations are 30 and 20 for the static and high frequency, respectively [**42**]. As in the case of parabolic bands, the screening effect is negligible at low densities, but here it becomes very strong at increased carrier densities, leading to a variation of more than an order of magnitude. The valence band of TiCoSb is strongly non-parabolic with wide energy surfaces. This leads to higher carrier density (and screening lengths) and especially wider exchange vectors, two factors which strongly increase carrier screening and increase scattering times and mfp (much more compared to the parabolic band cases).

*TDF illustrative examples:* We now show examples of transport distribution functions (TDF) $\Xi$, which contain all the information relevant to charge transport within the BTE. We first show examples of these functions for different scattering mechanisms in the case of parabolic bands. In **Figure 6a** we show $\Xi$ for the same parabolic band as in **Figures 5a-d**, for the three electron-phonon scattering mechanisms computed with *ElecTra*. As expected, ADP gives a linear $\Xi$ in the parabolic band case, while for ODP and POP the onset of the phonon emission at the characteristic phonon energy is clearly observed (here we used 50 meV for ODP and 40 meV for POP). Also, the effect of the anisotropic character of the POP scattering is noticeable, with higher than linear increase of $\Xi$, for the high energy, large exchange vector, carriers. In **Figure 6b** we compare the two $\Xi$ functions for ODP and POP from **Figure 6a** to the known analytical solutions, showing once again the validity and reliability of the *ElecTra* numerical implementation.

In **Figure 6c** we show the corresponding TDFs for the valence band of TiCoSb to indicate how a full-band treatment of non-ideal bandstructures impacts the transport distribution functions. $\Xi$ is no longer linear for ADP, and the abrupt jump at the phonon emission onset in ODP is mitigated (as observed in the logarithmic and linear-inset scales). Note that the assumed optical phonon frequency is 36 meV and that the figures need to be read from right to left. In **Figure 6d** we show $\Xi$ computed when considering all ADP, ODP and IIS altogether, and for three carrier densities, showing that an increasing dominant IIS tends to smear out some of the DOS features, although the more than linear trend expected from parabolic bands is absent – the trend here is rather linear. Note that the scattering parameters for TiCoSb are the same as used in ref. [**9**].

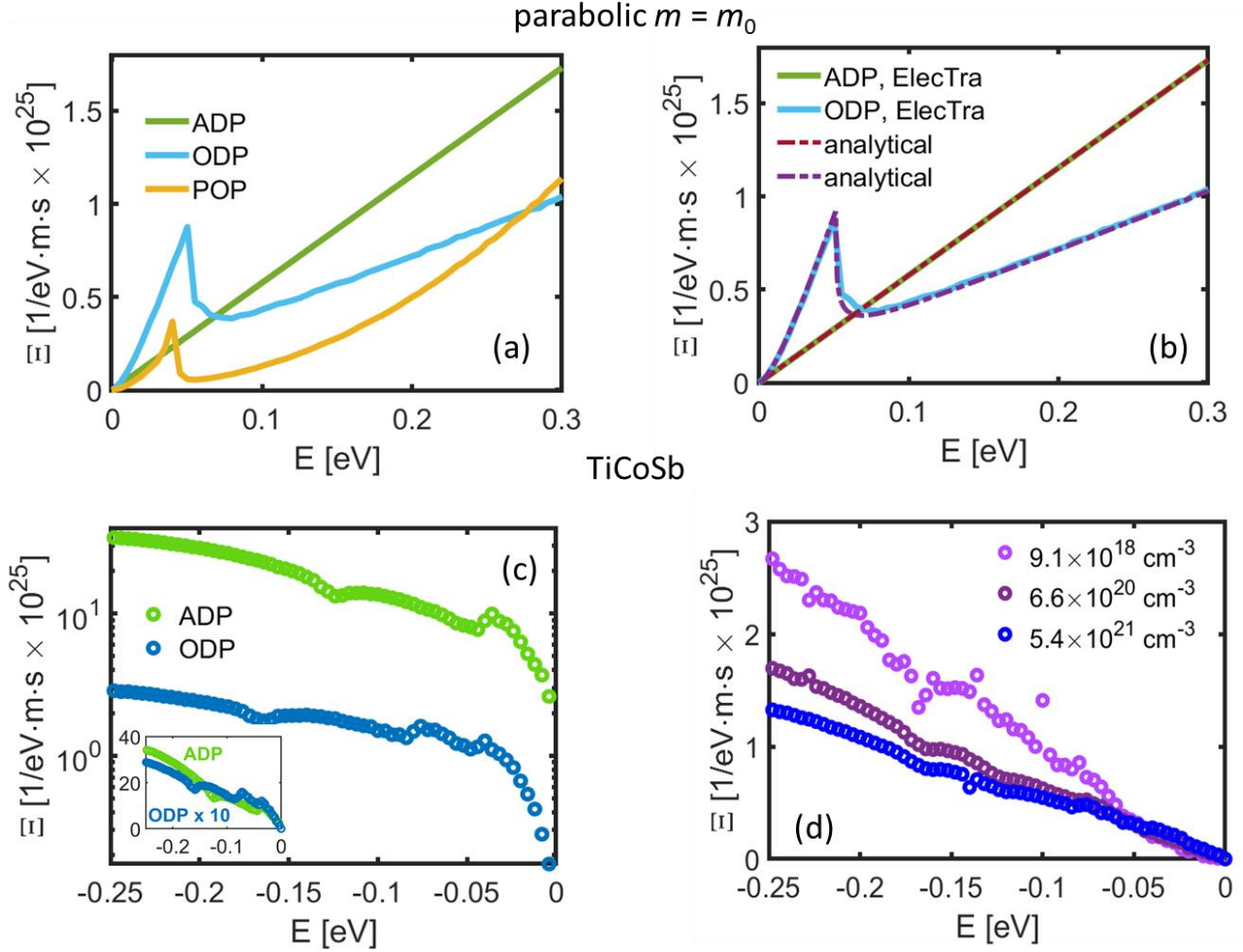

**Figure 6**: (a) The *xx* component of the TDF for an isotropic parabolic band for three different scattering mechanisms, as specified in the legend. (b) Validation of two TDFs in (a) with the analytical solution. (c) The *xx* component of the TDF for the TiCoSb valence band, for ADP and ODP mechanisms. The inset depicts the case for the *y*-axis in linear scale. (d) The total *xx* component of the TDF for TiCoSb for three different hole densities as specified in the legend. For TiCoSb, the shown data are for the 101×101×101 *k*-mesh.

## 5. Transport coefficient validations

Now we present the validation of the *ElecTra* simulator with comparison to BoltzTraP, a widely used simulation code that uses the constant relaxation time approximation (CRTA) **[12, 13]**. We also show simulations from *ElecTra* regarding the well-studied semiconductors Si, Ge, SiGe alloy, GaAs.

We start with a comparison between the TE coefficients computed under the CRTA by *ElecTra* and BoltzTraP for TiCoSb, i.e. the electrical conductivity, $\sigma$, in **Figure 7a** and the power factor, $PF = \sigma S^2$, in **Figure 7b**. A very good agreement is observed with a maximum difference of ~ 10%. In **Figure 7c** we report the comparison of $\sigma$ for two different tensorial components for another thermoelectric material, $Mg_3Sb_2$. Again the differences are below 10%. We also note that the mesh we used in this calculation is rather sparse, 51x51x51 for TiCoSb and 61x61x41 for $Mg_3Sb_2$.

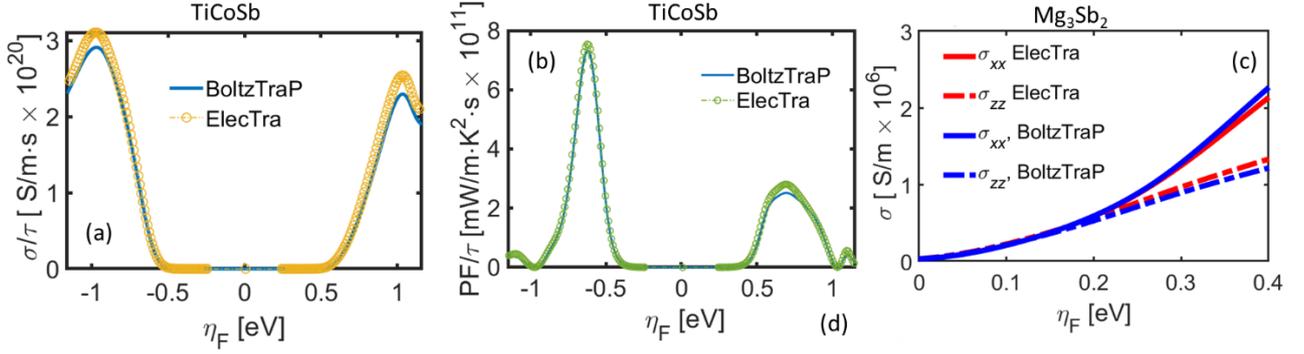

**Figure 7**: Validation of *ElecTra*, run under the CRTA, with BoltzTraP for TiCoSb, (a) and (b), and for Mg$_3$Sb$_2$, (c). The BoltzTrap results are in blue. The mesh used by *ElecTra* is 51×51×51 for TiCoSb and 61×61×41 for Mg$_3$Sb$_2$. These are in general sparser meshes than that we would have used, but for matter of comparison between the two codes they are adequate.

We now validate *ElecTra*'s transport calculations versus experimentally well-known semiconductors by comparing the computed and experimental mobility using data from the literature. [43-47]. In **Figures 8a-b** we show the case of *n*- and *p*-type silicon mobility respectively, including ADP, ODP and IIS with the figures, parameters, and full details of the calculation adopted from [**43**]. The very good agreement corroborates the validity of the implemented scattering and transport methods. The discrepancies at the higher doping region can be ascribed to electron-electron scattering, [**43**, **47**] which is not included in the computation.

GaAs offers the possibility to further validate the *ElecTra* scattering treatment by investigating a material for which POP is dominant. In addition, two type of valleys, Γ and L, participate in transport and different deformation potentials determine the scattering strengths between them [**22**]. GaAs also offers the possibility of testing the relevance of the carrier screening in the POP mechanism. As we observe in **Figure 8c**, the consideration of screening in the POP mechanisms has little effect, attributed to the small effective mass of the Γ valley, which results in larger screening length and smaller exchange vectors. As seen above, POP screening becomes important at higher carrier densities, but then the IIS is dominant. However, materials with large energy surfaces such as half-Heuslers could have a different behaviour. In **Figure 8c** we also show the mobility computed without the $\left(1 - \frac{v_{i(k\prime)}}{v_{i(k,n,E)}}\right)$ momentum relaxation term in Eq. (6), signalling its importance when the relevant scattering processes are anisotropic.

The case of Ge mobility is shown in **Figure 8d**, where again the *ElecTra* simulations map very well to experiments. This figure, however, also highlights the importance of the mesh spacing. Using units of π/*a*, where *a* is the lattice parameter, a spacing of 0.01 π/*a* corresponds to a 201×201×201 bandstructure mesh. A spacing of 0.02 π/*a* can be sufficient to capture the general trends in the transport properties and could be sufficient in materials screening/ranking or comparisons. However, for better quantitative accuracy, a spacing of maximum ~ 0.015 $\pi/a$ is recommended.

The case of the SiGe alloy, in **Figure 8e**, is used to validate the 'Alloy' scattering treatment. In this case, a low doping value of ~ $10^{14}$ cm$^{-3}$ is considered. For Ge, the scattering deformation potentials are taken from reference [**47**], and for the SiGe alloys, they have been linearly combined, weighted by the composition. Note that in the general case, there can be different number of processes for each material and a weighted deformation potential cannot be trivially extracted. For example, with regards to IVS in silicon, there are three tabulated IVS processes [**43**] while there are only two for germanium [**47**] (for example the Si g- and f-processes are not relevant for Ge). One first order approach (not unique) to tackle this would be to determine on an arbitrary basis the deformation potentials for the minimum number of processes that however provide the elemental material mobility. Because now both materials will be described by the same number of processes, we can combine these fictitious deformation potentials with Vegard's law. The calculation agrees very well with the experiments; however, we acknowledge a better process needs to be identified. Finally, in **Figure 8f** we compute the mean-free-paths for some of the materials, as indicated in the figure, for both non-degenerate (dash-dot lines) and heavily degenerate conditions (solid lines), indicating the capability of ElecTra to provide meaningful intrinsic transport quantities.

Recently, we have employed *ElecTra* to study the charge transport in Transparent Conductive Oxide (TCO) SnO$_2$, for which *ElecTra* simulations show very good agreement to experimental data for the mobility. [**48**]

We also highlight that *ElecTra* approach is suitable also for metals, semimetals, and gapless semiconductors. [**49, 50**] For zero and negative bandgaps, *ElecTra* can consider the scattering between the "valence" and "conduction" bands, which can even be the case of inelastic scattering in a semiconductor with a bandgap smaller than the phonon energy.

Finally, we make a note regarding the use of overlap integrals. This is usually considered to be unity [**51, 52**], which is valid for parabolic bands, but for other cases there are expressions that can also be used for different cases and materials. [**23**] The assumption of an overlap integral equal to the unity can lead to an overestimation of the scattering strength, since the overlap is generally less than unity. In ElecTra the user can choose to use for the generic valence band of a material the analytical expressions developed for the overlap integrals in the valence band of Si [**53**]. Note that when the deformation potentials are extracted from the matrix elements as in ref. [**26**], the overlap integrals and scattering selection rules are already included in those values.

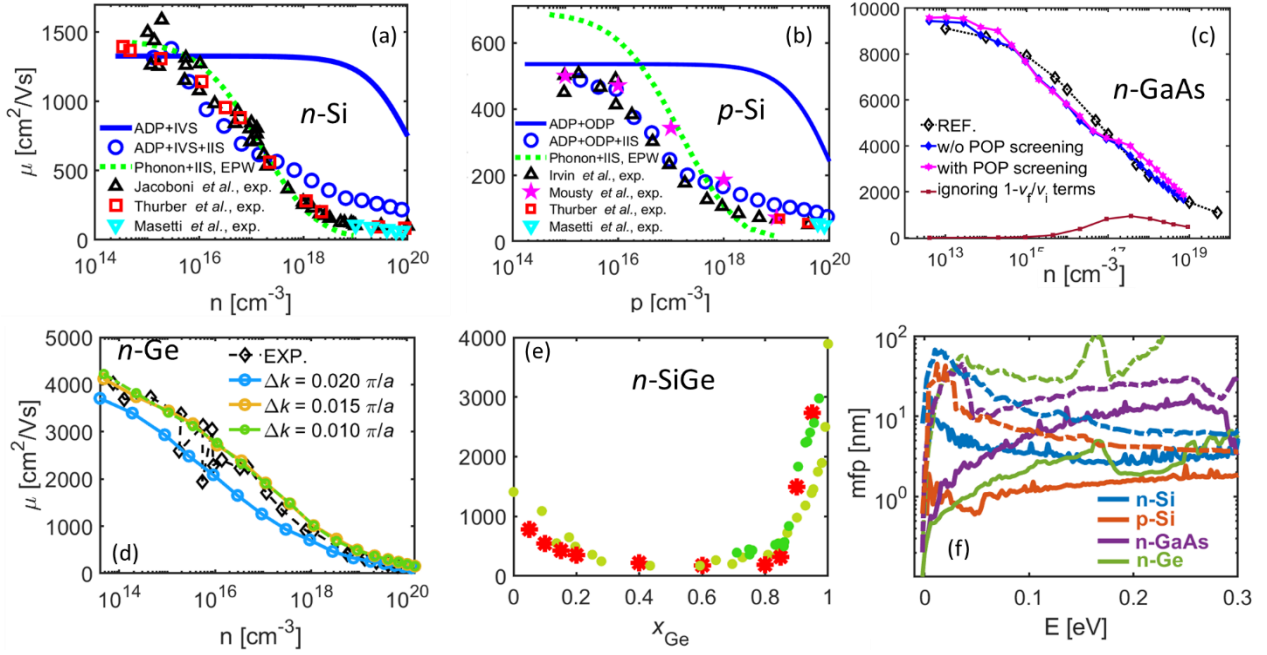

**Figure 8** (a-b) Simulation versus experiment for silicon electron and hole mobility versus carrier density at 300 K. The computation for Silicon uses DFT bandstructure (with SOC) and DFPT extracted deformation potentials [26]. (c) Simulation versus experiment for the electron mobility in GaAs; the conduction band has been numerically constructed from the nominal effective masses and non-parabolicity coefficients, so that we can model separately the two types of valleys. (d) Simulation versus experiment for the electron mobility in Ge. Nominal parameters for the bandstructure and scattering are used [22, 43-47]. The effect of the spacing in the *k*-space is displayed. (e) SiGe low doping electron mobility, the red stars are computed values whereas the green dots are experimental points. (f) mfps for (a)-(d) at low doping (dash-dot) with EF ~ 0.2 eV in the gap, and at high doping (solid) with EF ~ 0.1 eV into the band. "Reprinted (figures a-b) with permission from [Z. Li, P. Graziosi, N. Neophytou, Phys. Rev. B, 104, 195201 (2021)]. Copyright (2021) by the American Physical Society". [26].

## 6. Conclusions

We have introduced ElecTra, a code to solve the linearized Boltzmann transport equation by considering the full energy, momentum and band index dependence of the scattering relaxation time. The code has been tested versus analytical solutions, existing codes operating under constant relaxation time approximation, and experimental data, achieving an excellent agreement. The code can offer a large degree of accuracy with a significantly reduced computational cost compared to fully ab initio methods, and can be a useful resource in computing electronic and thermoelectric transport properties in complex bandstructure materials.

**Acknowledgement**

This work has received funding from the Marie Skłodowska-Curie Actions under the Grant agreement ID: 788465 (GENESIS - Generic semiclassical transport simulator for new generation thermoelectric materials) and from the European Research Council (ERC) under the European Union's Horizon 2020 Research and Innovation Programme (Grant Agreement No. 678763). We acknowledge support from Dr. Chathurangi Kumarasinghe in the validation of the algorithms for the formation of the constant energy surfaces, and form Prof. Laura De Sousa Oliveira for the initial set up of the parallelization concepts. For part of the computational time, we acknowledge the CINECA award under the ISCRA initiative, for the availability of high performance computing resources and support.

**Appendix A – 2D transport coefficients**

In this section we present the extension of *ElecTra* to 2D materials, report the scattering rate equations, and validate this computational method. In 2D the TDF is expressed as [**30**]:

$$\Xi_{ij(E,E_F,T)} = \frac{s}{(2\pi)^2}\frac{1}{t}\sum_{k,n}^{\mathfrak{C}_E^n} v_{i(k,n)} v_{j(k,n)} \tau_{i(k,n,E_F,T)} \frac{d\ell_{k_{\mathfrak{C}_E^n}}}{|\vec{v}_{(k,n)}|} \quad (A.1),$$

where $t$ is the material thickness, for example the unit cell size in the vertical direction, $\mathfrak{C}_E^n$ is the constant energy contour of band *n*, and $d\ell_{k_{\mathfrak{C}_E^n}}$ is the corresponding length element for the $k_{\mathfrak{C}_E^n}$ state, computed by reducing to two dimensions the approach described in Section 3; all other terms in Eq. (A.1) are the same as in Eq. (2). $\frac{d\ell_{k_{\mathfrak{C}_E^n}}}{|\vec{v}_{(k,n,E)}|}$ is the *k*-state density-of-states. It must be noted that in 2D the triangulation and nearest-neighbour approach requires nearly the same computation time for both the $k(E)$ extraction and scattering rates calculation, so the nearest-neighbour approach does not lead to any computational cost benefits in 2D. However, both options are still available to the user. For the parabolic band example below, the computational time to build the constant energy contours is around 10 s and for scattering rates and transport coefficients around 8 minutes, in a laptop computer with 6 processors, regardless the used sampling scheme, triangulation or nearest-neighbours.

The scattering rate calculations, Eqs. (7a), (7b), (7c), (7d), (7e), (7f), for all the mechanisms except POP and IIS, are generally the same, with the DOS of the final state expressed in terms of the constant energy contour length element. In Eqs. (7a), (7b), (7c), (7f), a 1/*t* multiplication (where *t* is a thickness in units of m) is required to account for the 2D nature of the system and the correct units, i.e. the mass density is still entered in kg/m$^3$ for ADP, ODP and IVS, and the 2D cell volume for 'Alloy' is still entered in m$^3$. Also, in the 'Alloy' scattering, the form factor of a hard sphere is substituted by the one of a hard disk. Thus, we don't reprint those equations here again. The POP and IIS scattering rates, on the other hand, have a different form and are computed as:

$$\left|S_{\boldsymbol{k},\boldsymbol{k}'}^{(POP)}\right| = \frac{\pi q_0^2 \omega}{2|\boldsymbol{k}-\boldsymbol{k}'|\varepsilon_0}\left(\frac{1}{k_\infty}-\frac{1}{k_s}\right)\left(N_{\omega,BE}+\frac{1}{2}\mp\frac{1}{2}\right)g_{\boldsymbol{k}'} \quad \text{(A.2a)}$$

$$\left|S_{\boldsymbol{k},\boldsymbol{k}'}^{(IIS)}\right| = \frac{\pi}{2\hbar}\frac{Z^2 q_0^4}{k_s^2 \varepsilon_0^2}\frac{N_{imp}}{|\boldsymbol{k}-\boldsymbol{k}'|^2+\frac{1}{L_D^2}}g_{\boldsymbol{k}'} \quad \text{(A.2b).}$$

The terms in Eqs. (A.2a), (A.2b) are the same as in Eqs. (7a), (7b), (7c), (7d), (7e), (7f) except $g_{\boldsymbol{k}'} = \frac{d\ell_{(\boldsymbol{k}',n,E)}}{|\vec{v}_{(\boldsymbol{k}',n,E)}|}$. The form of Eqs. (A.2a), (A.2b) is based on the fact that the Fourier transform (FT) [22, 23] of the Coulomb-like field $\sim 1/r$ in 2D is $2\pi/q$ instead of the 3D $4\pi/q^2$, and for a screened field the FT of $\sim$ the $1/r \exp(-L_D/r)$ is $\frac{2\pi}{\sqrt{q^2+L_D^2}}$ in 2D instead of the 3D $\frac{4\pi}{q^2+L_D^2}$, [54] where $r$ is the distance from the charge. Remarkably, the comparison between Eq. (A.2a) and Eq. (7d) reflects the expressions for 3D and 2D materials in the approximation where no dielectric is assumed to be surrounding the material and absence of image charges. [55-58] Similarly, the screening term which can be eventually added to the POP scattering mechanism becomes $\frac{1}{1+\frac{1}{|\boldsymbol{k}-\boldsymbol{k}'|^2}\frac{1}{L_D^2}}$. Finally, the screening length in 2D is computed as $L_D = 2\frac{k_s\varepsilon_0}{e}\left(\frac{\partial n}{\partial E_F}\right)^{-1}$, [59] where $n$ is the 2D carrier density.

The validation of 2D *ElecTra*'s scheme is represented in **Figure A1**, where the DOS and band velocity for an isotropic parabolic band with effective mass equal to the electron rest mass, are compared with the known exact solutions in (a) and (b), respectively. The two methods (triangulation and nearest neighbour sampling) from **Section 3** are considered as well. With regards to the scattering treatment, an exact solution for the ADP and ODP scattering cases can also be obtained. *ElecTra*'s results are comparable with the analytical calculations in (c) and (d) for the TDF and the electrical conductivity, respectively.

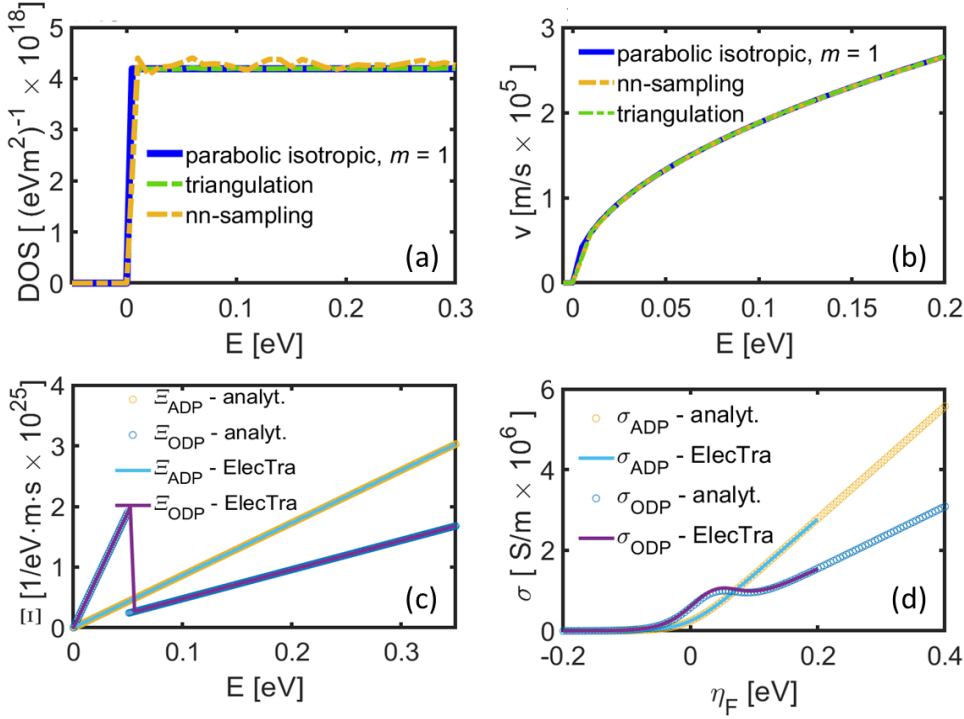

**Figure A1**: (a) Comparison of the DOS for a 2D parabolic band between the analytical value and the numerical one computed with *ElecTra*, by using the two BZ sampling techniques. (b) Same as in (a), for the band velocity comparison between the analytical solutions and numerical from *ElecTra*. (c) Comparison between the TDF computed with *ElecTra* with analytical known solutions for ADP and ODP scattering mechanisms. (d) Same as (c) for the electrical conductivity.

**Appendix B – additional technicalities**

In this appendix we provide some details of the code speed-up with parallelization and the available band interpolation. Additional information are provided in the Supporting Information and in the ElecTra manual. [**60**]

*ElecTra* is parallelized at the level of the carrier energy, and supports both local, multiple processors on the same node, and cluster, multiple processors on more nodes, parallelization. **Figure B1** shows the scaling performance, obtained on the CINECA Galielo100 cluster [**61**] on a single node, equipped with Intel CascadeLake 8260, each with 24 cores, of 2.4 GHz frequency, for bipolar calculations using TiNiSn considering ADP, ODP, POP (with screening) and IIS mechanisms, four temperatures and 15 doping levels (for each polarity). The speed-up shown is compared to the execution time using 5 CPUs as the basis, for (a) the constant energy surface composition, and (b) the rates and TDF calculations. The times are evaluated with the 'tic'-'toc' functions in MATLAB® and have around 5% of error. The times required to run the computations on 5 CPUs for the formation of constant energy surfaces are around 40 s and 70 s for conduction and valence bands respectively. For the scattering rates and transport coefficients calculations, they are 6.5 hr and 21.7 hr for the conduction and valence band, respectively. Fig. B.1d shows interpolation results for the DOS of the TiCoSb valence band. These numerical interpolations, which use the 'natural-neighbor' scheme as provided by a built-in MATLAB® routine, take around 15 minutes and 45 minutes for interpolating an initial 51 × mesh to a 101 × and 151 × meshes, respectively, on a laptop, which is similar time compared to the interpolation performed using Wannier interpolations, which can also be used as an alternative method.

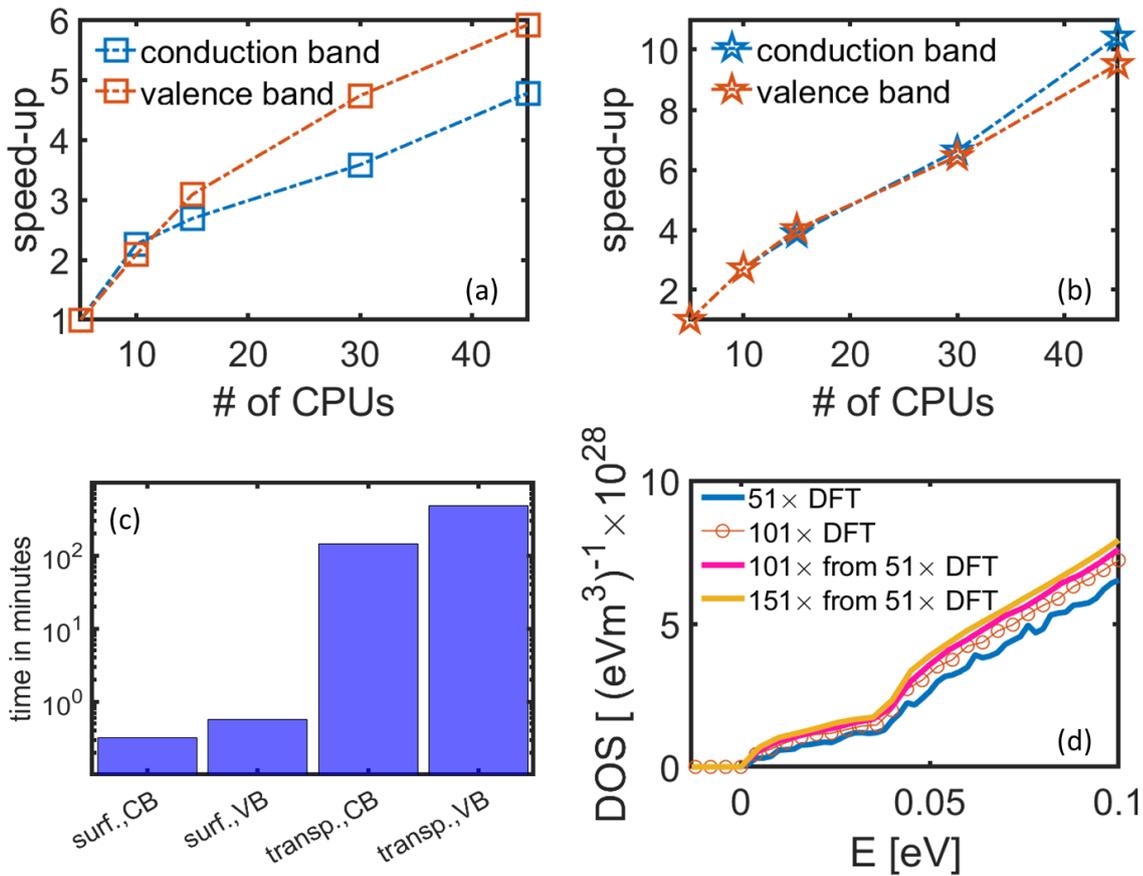

**Figure B1:** HPC speed-up in compared to a simulation of using 5 CPUs, for the case of TiNiSn and for: (a) the constant energy surfaces formation, and (b) the scattering times calculation. (c) Computation times for the case of 10 CPUs and the case detailed in the text as an example, for the computation of the constant energy surfaces, surf., and the scattering rates and transport coefficients, transp., for CB and VB, as indicated.(d) Numerical interpolation example for the TiCoSb DOS; solid lines are for a DFT mesh of 51×51×51 points and its interpolation, and compared to the results when using a 101×101×101 DFT mesh (red circles).

We address now discretization issues and how dense the *k*-mesh should be. We focus on the nearest neighbor (NN) sampling, whereas the comparison between NN and Delaunay triangulation (DT) schemes is displayed in **Figure 3f**. A convergence test for the DOS of the conduction band of Si is shown in Figure B2a, for different sets of ***k***-points extracted out of DFT. *ElecTra* can input in addition any interpolated DFT *k*-mesh, for example as provided by the Wannier90 package, and the accuracy will be the same as in any other transport code. *ElecTra* also offers a numerical interpolation of the *k*-space (using a built-in MATLAB® routine) to create a finer *k*-mesh and reach a larger number of points of what the bare DFT mesh would provide. We show the result in **Figure B2b** below, where we compare the result of the Si conduction band DOS for some numerical interpolations from an initial $30 \times 30 \times 30$ DFT mesh, together with the actual result using a $100 \times 100 \times 100$ DFT mesh. We then use the initial $30 \times 30 \times 30$ DFT mesh and we interpolate to $50 \times 50 \times 50$, $100 \times 100 \times 100$ and $150 \times 150 \times 150$ and use the NN sampling to extract the DOS. Good agreement is found to the $100 \times 100 \times 100$ DFT mesh when we interpolate the $30 \times 30 \times 30$ DFT mesh to $100 \times 100 \times 100$ at energies beyond 0.1 eV. To obtain good agreement at lower energies, however, as expected, an initial finer mesh must be used. Referring to **Figure B2**, it appears that a DFT calculation of a $50 \times 50 \times 50$ *k*-mesh, plus subsequent numerical interpolation to a $100 \times 100 \times 100$ mesh, would provide a reasonable trade-off between accuracy and DFT computational cost.

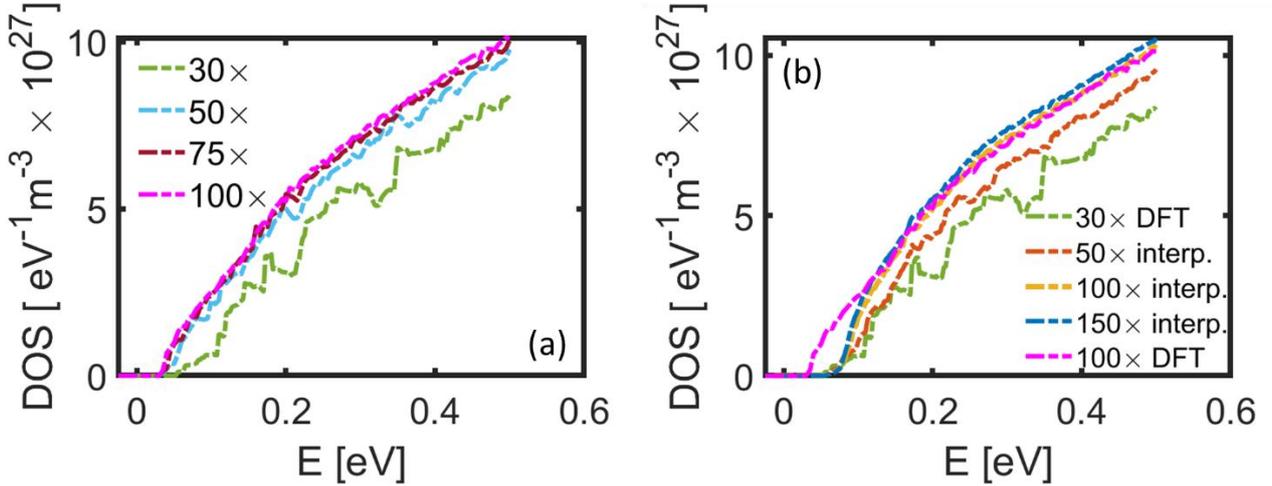

**Figure B2:** DOS for silicon conduction band (a) using different DFT *k*-meshes to sample the BZ using the NN sampling method. (b) starting from a 30×30×30 DFT mesh and interpolating it as indicated in the legend. The results are compared to the DOS results from a 100×100×100 DFT mesh. It appears that a mesh of at least 50×50×50, that corresponds to 0.035 $\pi/a_0$, must be used in the DFT calculation.

**Appendix C. Supplementary material**

The Supplementary material related to this article is listed after the references.

# Supplementary Material

Here we provide some details of how the code works with examples of interfaces, but a complete picture of the operation of the code can be found in the manual, and we refer the interested reader to that [1].

*GUI operation*: Below we show examples of the GUIs that can be used to operate *ElecTra* in this mode. Further detailed explanations can be found in the manual. *ElecTra* offers GUIs for entering the calculation instructions, preparing the scattering parameter file, and plotting the results. Below we describe these three GUIs in this order.

### a. *input app*

The calculation instructions can be entered in Basic or Advanced modes, depending on the user's confidence, knowledge of bandstructure details and needed output details. Two types of this app are available; one saves a .mat file to be passed to HPC clusters, while the other runs *ElecTra* directly. The second type is useful if the HPC supports GUI, or when executing *ElecTra* on a PC. **Figure B1** shows the tab to interface with the .bxsf data format, which can be used by many DFT codes, and the tabs to save the input files/run *ElecTra*. All the physical quantities are in International System units (SI), except the energy which is in eV.

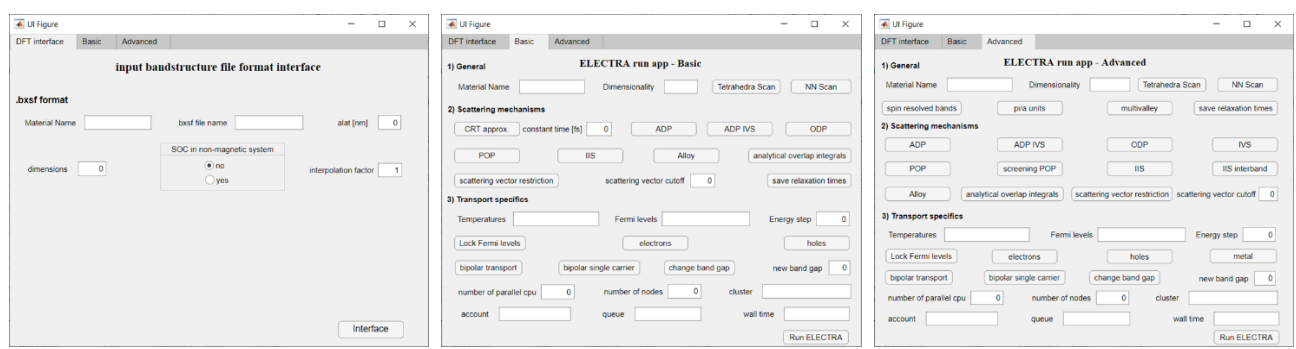

**Figure B1**: GUIs for the interface with bandstructure in .bxsf format, and the Basic and Advanced input mode. Many more details can be found in the code manual **[1]**.

The first tab of this GUI is for interfacing with DFT codes capable of saving the bandstructure in the .bxsf format. Here the user enters the material name, which must be the same one used later on for the transport calculation, the name of the .bxsf file, the value of 'alat', the lattice parameter used to express the reciprocal lattice vectors in the .bxsf file – this is generally an input value for DFT, but it is not saved in .bxsf files, the bands dimensionality, and if the bands are spin resolved but the material is non-magnetic; this is the case for SOC calculations for non-magnetic systems where single spin bands are returned by the DFT programs, but they are identical so that one of the two can be removed to speed-up the simulation. In this case, *ElecTra* considers the bands as double-spin and the button 'spin-resolved bands' must *not* be selected. In this interface the user can also choose to perform a numerical interpolation of the bands, based on the "natural neighbor interpolation" implemented in MATLAB®, which operates on each band separately. This is useful when a sufficient dense mesh cannot be achieved by DFT.

The second and third tabs are about the basic and advanced mode, respectively.
There are three levels of instructions:
*1) General information*
This includes the material's name, which is used to load the $E(\bm{k})$ file and the scattering parameters file, and the dimensionality of the system, that is specified as a number: 2 or 3. Then the user chooses the way the reciprocal space is scanned to build the constant energy surfaces. 'Tetrahedra Scan' adopts the tetrahedron methods, while 'NN Scan' performs the nearest-neighbour sampling. The 'NN Scan' scheme is developed to speed up the whole calculation. It is around 15× to 30× faster for 3D bandstructures. The computational speed of the two methods is, however, comparable in 2D. Only in the advanced mode, the user can use additional options such as specifications if the bands are spin-resolved as in a ferromagnetic material, if the $\bm{k}$-points positions are in units of pi/a (generally they are in 1/m units), or if the 'multivalley' option should be used. The latter then further allows in the scattering parameters file to use different parameters depending on the initial and final band-index pairs. *ElecTra* is parallelized in energy and can be operated in two ways: (i) the master CPU collects all the state-dependent relaxation times and after the parallel execution, composes the energy dependent quantities and integrates them, or (ii) the master CPU collects directly the energy dependent quantities so that the state individual relaxation times are used inside the CPU of the parallel pool and are not saved. Option (ii) is much more memory efficient. Option (i), although useful to look at very specific details, disentangle odd behaviours or debug the code, requires several tens of Gb to be accessible by the master CPU. This option, labelled 'save relaxation time' can be chosen in both the 'Basic' and the 'Advanced' operation tabs.
*2) Scattering mechanisms details*
The user can select the scattering mechanisms via the corresponding input fields. In the basic mode the user can also choose the constant relaxation time (CRT) approximation and input a constant relaxation time in femtoseconds. In this case, the code will also perform a constant mean-free-path (CMFP) approximation calculation using 5 nm as CMFP value (although the user needs to have in mind that this is in general not consistent with the CRT value). For sake of generality and flexibility, *ElecTra* doesn't require the state wavefunctions as inputs, hence the overlap integrals [2, 3] are taken to have values of one. However, depending on the user's choice, *ElecTra* can use an analytical form for the overlap integrals. This is usually considered to be unity [4, 5], but there are expressions that can also be used for different cases and materials [2]. In *ElecTra*, the user can choose to use, for the generic valence band material case, the analytical equations developed for the overlap integrals in the valence band of Silicon[6]. In the advanced interface, the user can also choose to add the screening to the POP or to make IIS also inter-band, while it is usually considered as intra-band only in the literature [7, 8, 9]. Finally, the user can choose to restrict the ADP and ADP-IVS scattering to only the *k*-states which are closer than a certain cut-off distance in the BZ, to reflect the fact that these scattering mechanisms assume linear acoustic phonon dispersions and negligible phonon energy and involve small wave-vector phonons. Such a distance is expressed as a portion of the BZ side length, for example if 0.2 is entered, the ADP will be allowed only if the initial and final points are distanced less than 0.2 times the BZ side length. In the Advanced mode, the user can also choose the 'IVS', Inter-Valley Scattering, mechanism, which describes the inter-band inelastic scattering. While in basic mode the ODP is both intra- and inter-band, in the Advanced mode the user can separate the two, as detailed in the user's manual [1].

3) *Transport specifics*

Here the user inputs the temperatures (in K) and the Fermi level positions (in eV) for which the simulation will be performed. In semiconductors, the Fermi level zero reference is automatically set at the majority carriers band edge, so negative Fermi level values place the Fermi level in the gap and positive values place it into the band. This is always the case for both band types, because when the user wants to simulate hole transport, the valence band is automatically flipped to positive energies. Thus, positive values of Fermi levels are always into the bands and negative values in the gap, regardless of conduction or valence bands simulations. For the case of full bipolar transport, in the current version v1, the same entered Fermi level values are used for both bands (each treated separately, so the user does not need to know the bandgap). *ElecTra* automatically finds the intrinsic Fermi level, adjusts the values inside the bandgap, and makes them symmetric for CB and VB. As each Fermi level corresponds to a specific doping level, it is preferable to input the Fermi level rather than the doping, because this gives the user a direct control over energy references. Indeed, in the generic material it might not be possible to know what doping levels correspond to degenerate or non-degenerate conditions.

In practice, the doping level in the material does not change with temperature (if the temperature is above the frozen region, which is always assumed by *ElecTra*) due to charge neutrality. Thus, by increasing the temperature, the Fermi distribution broadens, which forces the Fermi level to shift down away from the bands to keep the carrier density constant and equal to the doping density, as depicted in **Figure B2**. Here a parabolic band is sketched for two different impurity densities $N_i^{(1)}$ for non-degenerate doping in (a) and $N_i^{(2)}$ for degenerate doping in (b). The dashed lines represent the Fermi level positions at 300 K and 900 K. The Fermi distributions for these temperatures are depicted in (c) – blue line for 300 K and red line for 900 K. The broadening of the Fermi distribution at higher temperatures results in more states being occupied at higher energies. Thus, to keep the carrier density constant, the Fermi level shifts down. This shift is more pronounced for the non-degenerate conditions. Since the Fermi levels positions will correspond to different densities at different temperatures, we set that the *user entered Fermi levels* refer to the *user entered temperature closest to 300 K*. A unique carrier density and doping impurity density are found for each Fermi level. *ElecTra* assumes the semiconductor to be in the extrinsic region at that temperature, i.e. all the dopants are ionized and all mobile carriers come from the dopants. For a specific Fermi level, the impurity density (and carrier density) are used as a reference for the temperature closer to 300 K. For higher temperatures, *ElecTra* shifts the Fermi levels such that the impurity and carrier densities remain constant to their reference values at the entered temperature closer to 300 K. So, different sets of triads ($E_F$, charge density, T) are formed, and for the same carrier density we can have different $E_F$ depending on T.

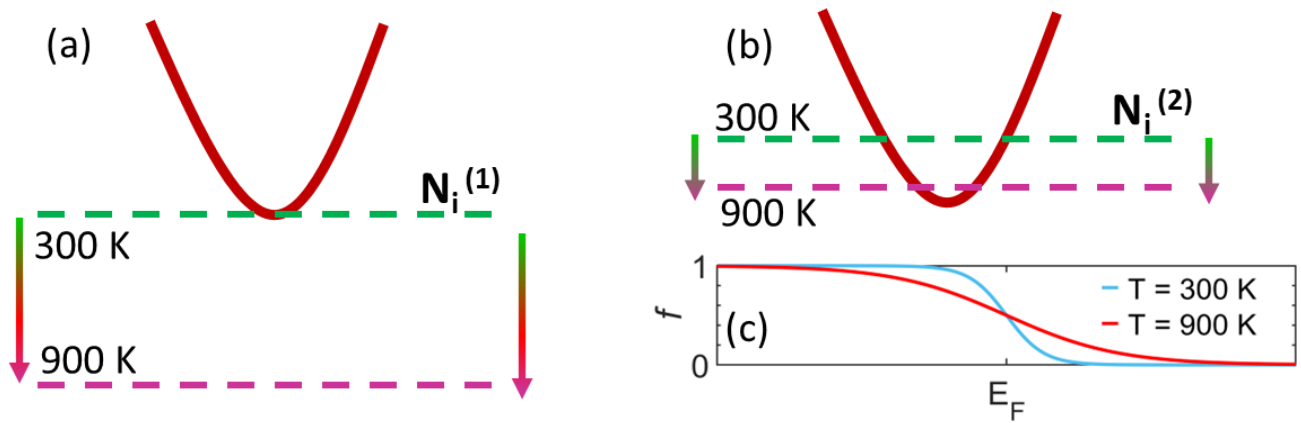

**Figure B2**: The shift in the Fermi level as the temperature increases from 300 K to 900 K for: (a) non-degenerate conditions, and (b) degenerate conditions. In (c), the Fermi distributions are shown for the two temperatures.

However, a user might want to fix the Fermi level positions for all temperatures, being aware that these will correspond to different doping/carrier densities. In this case the user can "lock" the Fermi levels by selecting the 'Lock Fermi levels' button, so that the entered Fermi levels will be used at all temperatures without considering $E_F$ shifts (i.e. for a certain $E_F$ we can have different charge densities depending on T).

Bipolar transport, [10, 11] can be simulated by selecting the 'bipolar transport' entry. When also the 'bipolar single carrier' is selected, *ElecTra* performs the calculation by placing the Fermi levels only around the selected majority carriers (but still considering the presence of the minority band when it comes to computing the charge density and IIS rates). This reproduces an experimental situation where a specimen is *n*-doped or *p*-doped. When this button is NOT selected, the code performs a full bipolar calculation (scans with $E_F$ both bands), i.e. the code will compute the transport properties for the *n*-type case, electron majority and hole minority, as well as *p*-type case, hole majority and electron minority. In the case of bipolar transport, the user can choose to change the band gap and enter a desired bandgap value. This can obviate the difficulty in computing reliable band gap values within DFT and generally offers more flexibility to the user. The new band gap must be entered with a value in eV.

If the material system does not have a bandgap, the 'metal' option can be selected. In this case the distinction between the conduction and valence bands is not made. Semimetals and gapless semiconductors, [12, 13] worth a special mention. If the bandgap is zero, they can be simulated as full bipolar cases as well as metals, but if it is negative, must be considered as a metal. In the case where we want to allow the scattering between the valence and conduction bands, which can even be the case of inelastic scattering in a semiconductor with a bandgap smaller than the phonon energy, the material must be considered as a metal; *ElecTra* will consider the polarity of the carrier using the Seebeck sign.

Finally, the parallelization details are selected. The entry 'number of parallel cpu' is the CPU number on the node on which the code will run, and this must be coordinated with the specific HPC cluster facilities. For multimode cluster parallelization other information must be introduced, which depend on the specific cluster. The number of nodes can be set to zero or one if a single node parallelization is desired. The other fields generally used in multimode cluster parallelization can be left empty if not needed.

    **b.** *<u>scattering app</u>*

The second GUI contributes to the scattering file preparation and is shown in **Figure B3**.

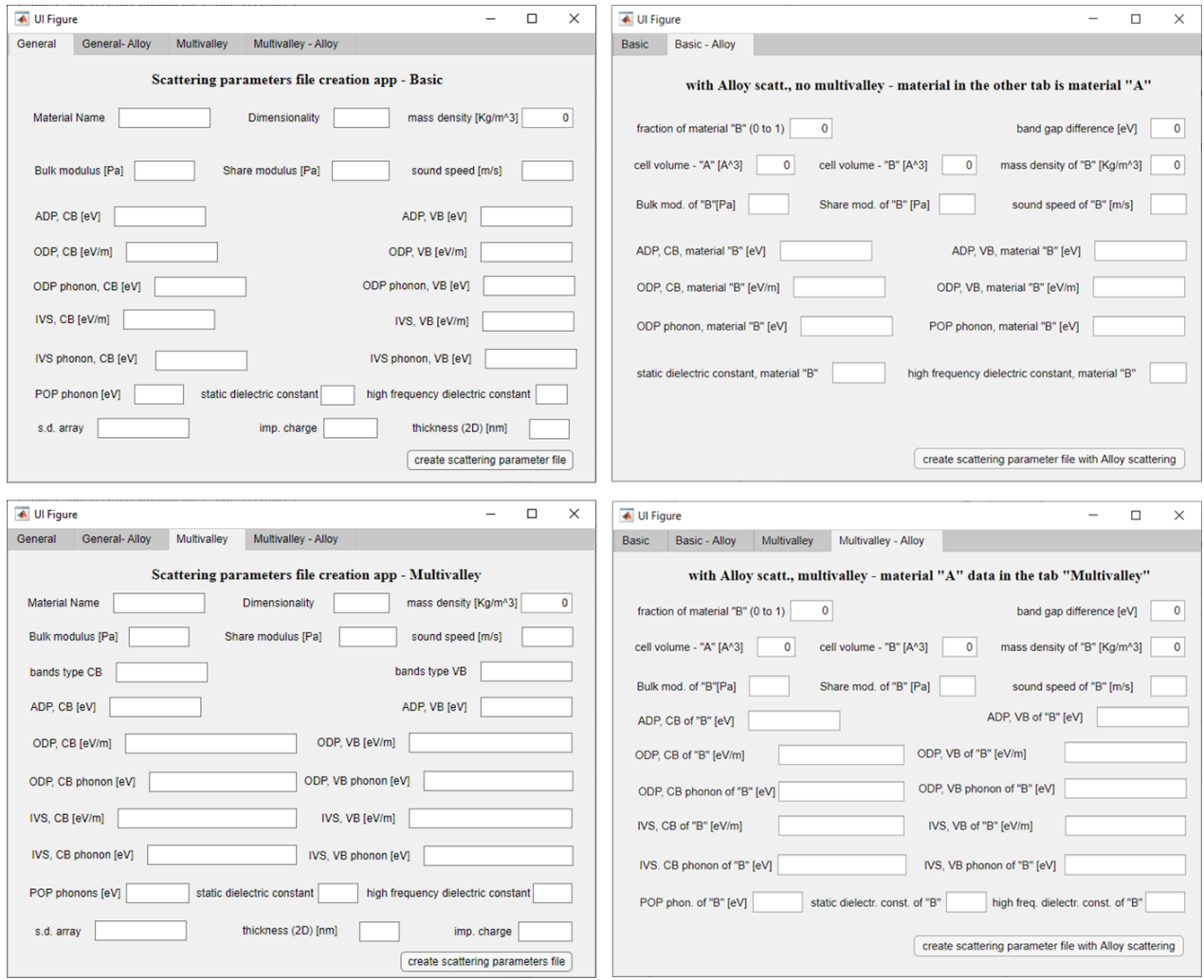

**Figure B3**: GUIs tabs for the preparation of the scattering parameters file, from left to right we show the case without Alloy scattering, the case with alloy scattering, and the case for the multivalley operation.

The first tab is for the standard file and contains the material parameters required for the calculation of the scattering rates, i.e. Eqs. 7 except the alloy scattering. The speed of sound can be inserted in two ways: the bulk modulus *K* and the shear modulus *G* can be inserted to compute the speed of sound, [15] *alternatively*, the speed of sound can be directly inserted; in that case *K* and *G* will be ignored. In the case of the alloy scattering between two 'A' and 'B' parental compounds, the parameters in the first tab refer to the parental compound 'A' and the additional required parameters, together with the ones for the parental compound 'B', are entered in the second tab. The third and fourth tabs correspond to the first two tabs for the multivalley mode, which allows the user to use different scattering parameters for different pairs of initial and final bands. Each band must be labelled by a number in the 'band_type' entry and each scattering parameter will be represented by a matrix whose element *i,j* corresponds to the relative scattering transition between bands *i* and *j*. Several examples on how this can be done are shown in the manual. [**1**]

### c. *plot app*

*ElecTra* results can be plotted using a dedicated GUI. The app plots the selected quantities in a graph inside the app and in an independent figure. In addition, the app saves the plotted data in a .csv file in the app directory. The .csv file is designed to be opened with programs like Origin® and contains two or more columns where the first is the x-axis variable and the other(s) is (are) the y-axis quantities. The header contains the column name and specific additional description.

The app composes of two tabs, the first is designed to plot the charge transport coefficients whereas the second to plot the energy dependent quantities. These are shown in **Figure B4**.

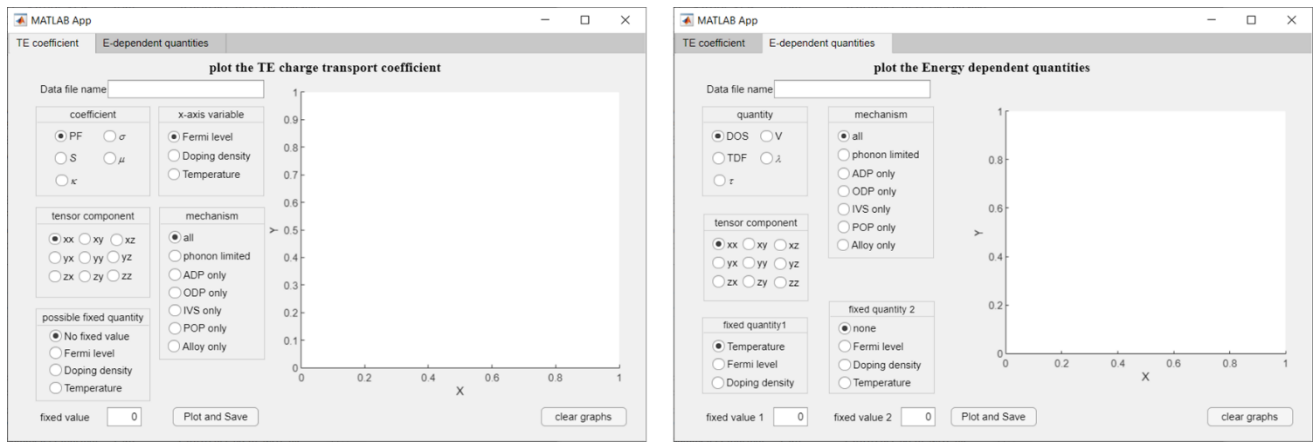

**Figure B4**: GUIs to plot the simulated results, the TE coefficients (left) and the energy-dependent quantities (right). The data are plotted in the GUI graph and in a separated figure, and are saved as .csv files.

The first tab, devised to plot the thermoelectric coefficients, 'TE coefficient', allows the user to enter first the file name to be opened, which contains the *ElecTra* data. Then, the user selects the transport coefficients to be plotted: PF (Power Factor), $\sigma$ (electrical conductivity), *S* (Seebeck coefficient), $\mu$ (mobility), $\kappa$ (electronic thermal conductivity); this will be displayed in the y-axis. The user also chooses the desired tensor component and the x-axis variable. After this, the user chooses if displaying the overall specific transport coefficient or only the contribution from one transport mechanism, e.g., what is the PF if only ADP is considered. When the x-axis is the doping density and the transport simulation is a full bipolar case, the carrier density for *p*-type is displayed as a negative value.

Finally, we note that a transport coefficient usually depends on more than one quantity; it usually depends on both Fermi level position (carrier concentration) and temperature. Thus, if the Fermi level is selected as the x-axis variable, a transport coefficient for each temperature used in the simulation will be displayed. If the user wants to display the value for only one temperature, the 'desired fixed value' can be selected, in this case 'Temperature', and the temperature of interest can be entered in the 'fixed value' space. Alternatively, if the x-axis is the temperature, a single value of Fermi level or carrier density can be chosen. The app will plot the data for the fixed calculated value which resides closer to the requested one. *NOTE*: in the 'fixed value' field, the Fermi level is in eV, the Doping density in $cm^{-3}$, and the temperature in K.

The "Plot and Save" button will plot the data in the app graph and in an independent figure, and save the .csv file which will have the name of the transport coefficients, e.g., 'PF_xx.csv'. The "clear graphs" button clears both the graphs.

The second tab is used to plot the energy dependent quantities, 'E-dependent quantities'. The user first enters the file name and chooses the quantity to plot; here these are the DOS (density of states), V (band velocity), TDF (transport distribution function, eq. 2), $\lambda$ (mean-free-path, eq. 6) and $\tau$ (relaxation time, eq. 7). The tensor component is selected as in the previous tab (i.e. xx, yy, xy ...). For $\lambda$ and $\tau$, which do not have mixed components, only the first letter of the tensor component is used, and for DOS and V no tensor component is considered. The scattering mechanism is selected as in the previous tab.

In this tab the x-axis is always the carrier energy, so some of these quantities have three variables ($E$, $E_F$, $T$). For graphical reasons, at least one value must be fixed, which can be the Temperature or the Fermi level (Doping density), and has to be selected in the "fixed quantity 1" menu and defined in the field below. As above, the app will plot the quantities for the fixed calculated values closer to the one entered (which might not be the same as the calculated ones). The units are as above: eV for Fermi level, $cm^{-3}$ for doping density, K for temperature. Eventually the user can opt for a second fixed quantity, different from the first. In this way the user will see only one line for the chosen temperature and Fermi level (doping density) values. The DOS and V ('complete' band velocity data) do not require any fixed value as these are bandstructure quantities.

The "Plot and Save" and "clear graphs" buttons work as in the TE coefficients tab.

***Text files operation***:

Here we show template examples of the input and scattering files. A user used to run calculations in HPC infrastructure without graphical interface support can find these more suitable. The user must beware that the GUIs save the data as *.mat* file whereas the text files have a *.m* extension; *ElecTra* will automatically check for both and *will prioritize the .m file*.

### a. input file

Below we show an example of an input file for *ElecTra*. It must be saved in the working directory and must be named 'input_file_ELECTRA_v1.m' (with the 'v1' standing for version 1). The file here is filled for a bipolar calculation for TiCoSb. The file structure maps the one described in the GUI, where for each GUI button there is a yes/no instruction, detailed with comments, and with the numerical fields being arrays.

```
% 1) GENERAL SETTINGS
material_name = 'TiCoSb';
file_type = 'bxsf'; % bxsf % for .bxsf files
bxsf_file_name = 'TiCoSb_fs.bxsf';
alat = 0.417 ; % in nm
SOC = 'no'; % yes or no
interpolation_factor = 2;

dimensionality='3D'; % 2D, 3D

scan_type='kScan'; % kScan or DT. This is the way the k(E) is extracted

spin_resolved = 'no';

k_units_SI='yes'; % in 1/m , as extracted from bxsf, okay for DFT bands
k_units_pi_a='no'; % usable for numerically built bands

%------------------------------------------------------------------------
%------------------------------------------------------------------------

% 2) SCATTERING MECHANISMS
ADP = 'no'; % Acoustic Deformation Potential, INTRAvalley
ADP_IVS = 'no';   % INTRA- and INTER- valley
k_restriction = 'no'; % only for the ADP_IVS TO BE EXTENDED WHEN VALIDATED
k_restriction_cutoff = 0.2; % Portion of the BZ to be considered

IVS = 'no'; % intervalley scattering

ODP = 'no'; % Optical deformation potential

POP = 'no'; % polar optical phonon, constant freq., approx.
screening_POP = 'no';

IIS = 'no'; %
IIS_interband_flag = 'no';

Alloy = 'no';

overlap_integrals_analytical = 'no';
```

```matlab
save_RTstruct = 'no';

multivalley = 'no'; % different types of valleys with different Deformation
Potentials

constant_tau = 'no'; % constant relax. time approx. (CRTA)
tau_const = 1e-14; % in s; value of the scattering time in the CRTA

%------------------------------------------------------------------------
%------------------------------------------------------------------------

% 3) TRANSPORT CONDITIONS

% 3.1) FERMI LEVELS AND TEMPERATURES
EF_array = [-0.2,-0.1,-0.06:0.02:0.14,0.2];   % in eV
% Fermi array is in respect to the band edge that will be set to zero,
% negative values mean Fermi into the gap, positive, into the band (degen.)

T_array = 300 :100: 900; % in K
Fermi_shift_flag = 'yes'; % to shift the Fermi level with the temperature

% 3.2) CARRIERS AND ENERGY
carriers = 'electrons'; % 'holes' or 'electrons'

bipolar_transport = 'yes';

bipolar_single_carrier = 'no' ;

change_band_gap = 'no';
new_band_gap = 1 ; % in eV

metallic = 'no'; % proper metal with Fermi in the bands free carriers without
impurities

Estep = 0.002; % energy step in eV, 2 meV is often largely sufficient

%------------------------------------------------------------------------
%------------------------------------------------------------------------

% 4) PARALLELIZATION
type_of_parallelization = 'cluster'; % 'local' (one node) or 'cluster' (more
than one node), 'MATLAB Parallel Cloud' is forecasted
max_number_of_cpu = 15 ; % maximum number of cpu in the single node
% only for cluster type
number_of_nodes = 2; % number of requested nodes
cluster_name = 'galileo100 R2020b';
account_name = 'IscrC_sd-FRAME';
queue_name = 'g100_usr_prod';
wall_time = '03:40:00';
```

**b. *scattering file***

The scattering parameters files are must be named 'scattering_parameters_'material'.m', e.g. scattering_parameters_TiCoSb.m , and must be saved in the "Data" folder. Examples of the nomenclature are as follows. If we consider for instance the variable D_adp_e, 'D' stands for the deformation potential, 'adp' for ADP, and 'e' for electrons (conduction band). The phonon frequency is called, for instance, hbar_w_odp_h, where 'hbar_w' is for $\hbar\omega$, 'odp' for ODP, and 'h' for holes (valence band), and so on. The static and high frequency dielectric constants, in units of vacuum permittivity, are named k_s and k_inf respectively.

```
rho_mass_density = 7.42*1e-3/(1e-2)^3; % kg/m^3
materialsproject.org/materials/mp-5967/
K_bulk_modulus = 142e9; % Pa
G_shear_modulus = 76e9; % Pa
sl = sqrt((K_bulk_modulus+4/3*G_shear_modulus)/rho_mass_density);  % m/s ,
elasticity theory
st = sqrt(G_shear_modulus/rho_mass_density);                       % m/s , elasticity
theory
us_sound_vel = 1/3*sl+2/3*st;                  % m/s   Ottaviani 1975
% experimental sl = 5.69e3; st = 3.23e3;

% ------- for holes ------------------
D_adp_h = 0.5; % eV, DOI: 10.1038/s41467-018-03866-w
%
D_odp_h = 2.20e10; % eV/m, DOI: 10.1038/s41467-018-03866-w, multiplied by 1/5 of
theBZ

% ------- for electrons ------------------

D_adp_e =1; % eV, DOI: 10.1038/s41467-018-03866-w

D_odp_e = 1.75e10; % eV/m, DOI: 10.1038/s41467-018-03866-w, multiplied by 1/5 of
theBZ

hbar_w_odp = 0.036 ; %eV, materialsproject.org/materials/mp-5967/  from the
centre of the LO  phonons DOS
Z_f_adp = 1; % number of final available valleys
Z_f_odp = 1; % to be defined as a row for every scattering mechanism, in any row
an element for each available band

hbar_w_pop=0.038; % eV, materialsproject.org/materials/mp-5967/ value in Gamma

% ------- Coulomb scattering - screened -------
k_s = 19.09; % from Zhou (DOI: 10.1038/s41467-018-03866-w) private communication
k_inf = 5; % F.G. Aliev et al. Zh. Eksp. Teor. Fiz. 47 (t988) t5l
Z_i = 1; % ionized impurity charge
```